\documentclass[12pt,a4paper]{article}
\usepackage{amsmath,amsthm,amsfonts,amssymb,cite}
\usepackage[dvips]{graphicx}
\usepackage[cp1251,ctt]{inputenc}
\usepackage[english]{babel}
\usepackage[T2A]{fontenc}

\DeclareSymbolFont{bletters}{OML}{cmm}{bx}{it}
\DeclareMathSymbol{\bla}{\mathord}{bletters}{'025}
\DeclareMathSymbol{\bmu}{\mathord}{bletters}{'026}
\DeclareMathSymbol{\bth}{\mathord}{bletters}{'022}
\DeclareMathSymbol{\bfI}{\mathord}{bletters}{"49}
\DeclareMathSymbol{\bdl}{\mathord}{bletters}{"0E}
\DeclareMathSymbol{\bDl}{\mathord}{bletters}{"001}
\def \bpi{\boldsymbol\pi}
\def \bphi{\boldsymbol\phi}

\def \Ga{\Gamma}
\def \si{\sigma}
\def \la{\lambda}
\def \be{\beta}

\def \ta{\theta}
\def \dl{\delta}

\def \CA{\mathcal A}
\def \CK{\mathcal M}

\def \CN{\mathcal N}
\def \CV{\mathcal V}

\def \CP{\mathcal P}

\def \CT{\mathcal T}
\def \BC{\mathbb{C}}
\def \BN{\mathbb{N}}

\def \BZ{\mathbb{Z}}
\def \BI{\mathbb{I}}

\def \d{{\rm{d}}}

\begin{document}
\title{
$${}$$
{\bf Correlation functions of XX0
Heisenberg chain, $q$-binomial determinants,
and random walks}}
\author{
\\
{\bf\Large N.~M.~Bogoliubov,
C.~Malyshev}\\[0.5cm]
{\it\small St.-Petersburg Department of Steklov Mathematical Institute RAS}\\
{\it\small Fontanka 27, St.-Petersburg,
191023, Russia} }

\date{}

\maketitle

\begin{abstract} \noindent
The $XX0$ Heisenberg model on a cyclic chain is considered. The representation of the Bethe wave functions via the Schur functions allows to apply the well-developed theory of the symmetric functions to the calculation of the thermal correlation functions. The determinantal expressions of the form-factors and of the thermal correlation functions are obtained. The $q$-binomial determinants enable the connection of the form-factors with the generating functions both of boxed plane partitions and of self-avoiding lattice paths. The asymptotical behavior of the thermal correlation functions is studied in the limit of low temperature provided that the characteristic parameters of the system are large enough.
\end{abstract}


\thispagestyle{empty}

\newpage

\section{Introduction}
\label{cor:sec0}

The exactly solvable Heisenberg $XXZ$ model is a prominent model describing the interaction of spins $\it{\frac12}$ on a chain. The integrability of the model via the algebraic Bethe Ansatz has led to important results, going from the spin dynamics up to the exact expressions for the correlation functions \cite{ft, kuls, KBI1, KBI2, vk, korr, vk1, ess, ml2, ml5, ml4}.

The $XX0$ Heisenberg chain is the zero anisotropy limit of the
$XXZ$ model, it also may be considered as a special free fermion case of the $XY$ magnet \cite{lieb, its}.
It appears that $XX0$ model is related to many mathematical problems. It is related to the theory of the symmetric functions \cite{macd} and to the theory of plane partitions. Plane partitions (three-dimensional Young diagrams) \cite{macd, andr, bres} were then discovered to be connected with amazingly wide ranging problems in mathematics as well as theoretical physics. They are intensively studied, e.g., in probability theory \cite{vers, borod}, enumerative combinatorics \cite{stan}, theory of faceted crystals \cite{ferr, oko1}, directed percolation \cite{raj},
topological string theory \cite{oko2}, and the theory of random walks on lattices \cite{bres, 5, b1, b2}.

The correlation functions of the $XX0$ chain are of considerable interest, and their behavior was intensively investigated for the system in the thermodynamic limit \cite{ml2, iz11, vk3, iz5}. In our paper we study
the asymptotical behavior of the thermal correlation functions in the limit of low temperature provided that the chain is long enough while the number of flipped spins is moderate. Namely in this limit the thermal correlation functions are related to random matrix models \cite{b1}. This connection allows to uncover, in particular, the mapping between the correlation functions and the low energy sector of quantum chromodynamics \cite{iz5}.

We shall consider the $XX0$ Heisenberg model on the periodical chain. The representation of the Bethe wave functions via the Schur functions \cite{macd} allows to apply the well-developed theory of the symmetric functions to the calculation of the thermal correlation functions as well as of the form-factors. In the present paper we are interested in the correlation functions of two types: the correlation function of the states with no excitations on $n$ consecutive sites of the chain that will
be called {\textit {persistence of
ferromagnetic string}}, and the
correlation function of the creation operator of the $n$ excitations on the consecutive sites of the
chain that will be called {\textit
{persistence of domain wall}}. Special attention will be paid to the combinatorial objects appearing
in the calculations (the generating functions of plane partitions and random walks, the
$q$-binomial determinants) and to
the combinatorial interpretation of the obtained results. We will calculate the leading terms of their asymptotics, provided that the characteristic parameters of the system are large enough, including the critical exponents of these correlation functions in the low temperature limit, and the related amplitudes. These amplitudes are found to be proportional to the squared numbers of boxed plane partitions.

The paper is structured as follows.

Section~\ref{cor:sec0} is introductory.
The $XX0$ model and its
solution are presented shortly in Section~\ref{cor:sec1}, the considered correlation functions are defined and the amplitudes of the state vectors are written in terms of Schur functions. This representation allows to calculate the form-factors of operators in  Section~\ref{cor:sec2} applying the
formulas of the Binet-Cauchy type. The persistence of ferromagnetic string as well as the persistence of domain wall are also calculated in this section.
In Section~\ref{cor:sec3} we deal with the combinatorial aspects of the problem. The $q$-binomial determinants are introduced and their connection with the generating functions of plane partitions is discussed. It is shown also that the form-factors, obtained in the previous section, under the special parametrization are expressed as the generating
functions of boxed plane partitions and of the self-avoiding lattice paths.
The asymptotical estimates of the correlation functions are obtained in Section~\ref{cor:sec5}. Discussion in Section~\ref{cor:sec6} concludes the paper. In Appendix~I we provide some notions concerning boxed plane partitions and their generating functions.  The proof of the determinantal formulas crucial for this paper is given in Appendix~II.

\section{XX0 Heisenberg model and outline of the problem}
\label{cor:sec1}

The Heisenberg $XX0$ model  on the chain of $M+1$ sites is defined by the Hamiltonian
\begin{equation}
{\cal H} \equiv - \frac 12\sum_{k=0}^M
(\si_{k+1}^{-}\si_k^{+} +
\si_{k+1}^{+}\si_k^{-})\,. \label{anis1}
\end{equation}
Here the periodic boundary conditions $\si^{\#}_{k+(M+1)}=\si^{\#}_k$ are assumed. The local spin operators $\si^\pm_k = \frac12 (\si^x_k\pm i\si^y_k)$ and $\si^z_k$ obey the commutation rules:
$[\,\si^+_k, \si^-_l\,]\,=\,\dl_{k
l}\,\si^z_l$, $
[\,\si^z_k,\si^\pm_l\,]\,=\,\pm 2\,\dl_{k
l}\,\si^{\pm}_l$ ($\dl_{k l}$ is the
Kronecker symbol). The spin operators act in the space ${\mathfrak H}_{M+1}$ spanned over the states $\bigotimes_{k=0}^M
\mid\!\! s\rangle_{k}$, where $\mid\!\!
s\rangle_{k}$ implies either spin ``up'', $\mid\uparrow\rangle$, or spin ``down'',
$\mid\downarrow\rangle$, state at k$^{\rm th}$ site. The states $\mid\uparrow\rangle\equiv
\begin{pmatrix}
1 \\
0
\end{pmatrix}$ and $\mid\downarrow\rangle
\equiv
\begin{pmatrix}
0 \\
1
\end{pmatrix}$ provide a natural basis of the linear space ${\BC}^2$.

The sites with spin
``down'' states are labeled by the
coordinates $\mu_i$, $1\leq i\leq N$. These coordinates constitute a strictly decreasing partition ${\bmu}=(\mu_1, \mu_2,\,\dots\,, \mu_N)$, where $M\geq\mu_1>\mu_2>\,\dots\,>
\mu_N \geq 0$. The other important partition is ${\bla}=(\la_1, \la_2, \dots, \la_N)$ of
weakly decreasing non-negative integers:
$L\geq\la_1\ge\la_2\ge\,\dots\,\ge \la_N\geq 0$. The elements $\lambda_j$ are called the \textit{parts} of $\bla$. The \textit{length} of partition  $l(\bla)$ is equal to the number of its parts. The sum of all parts is the \textit{weight} of partition, $|\bla|=\sum_{i=1}^N\la_i$.
Partitions $\bla$ can be represented by Young diagrams. The Young diagram of $\bla$ consists of $N$ rows of boxes aligned on the left, such that the $i^{\rm th}$ row is right on the $(i+1)^{\rm st}$ row. The length of the $i^{\rm th}$ row is $\la_i$.
The relation
$\la_j=\mu_j-N+j$, where $1\le j\le N$, connects the parts of $\bla$ to those of $\bmu$. Therefore, we can write: $\bla =\bmu - {\bdl}_N$, where ${\bdl}_N$ is the strict
partition $(N-1, N-2, \dots, 1, 0)$.
There is a natural correspondence between the coordinates of the spin
``down'' states $\bmu$ and the partition $\bla$  expressed by the Young diagram (see Fig.~1). Throughout the paper bold-faced letters are used as short-hand notations for appropriate $N$-tuples of
numbers.
\begin{figure} [h]
\center\includegraphics {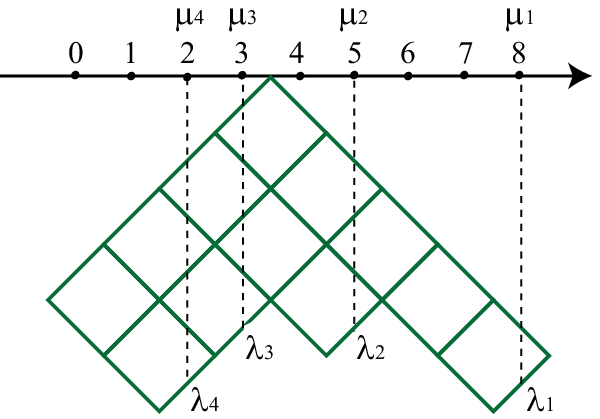}
\caption{Relation of the spin ``down'' coordinates $\bmu=(8, 5, 3, 2)$ and partition $\bla=(5, 3, 2, 2)$ for $M=8$, $N=4$.}
\end{figure}

The $N$-particle state-vectors  $\mid\!\Psi({\bf u}_N)\rangle$, the states with $N$ spins ``down'', is convenient to express by means of the Schur functions \cite{b4, b5}:
\begin{equation}
\mid\!\Psi({\textbf u}_N)\rangle =
\sum\limits_{\bla \subseteq \{{\CK}^N\}}
S_\bla ({\textbf u}^{2}_N) \begin{pmatrix}
\prod\limits_{k=1}^N \si_{\mu_k}^{-}\end{pmatrix} \mid
\Uparrow \rangle\,, \label{conwf1}
\end{equation}
where summation is over all partitions
$\bla$ satisfying $\CK\equiv M+1-N\geq \la_1\geq \la_2\geq \dots\geq
\la_N\geq 0$. The parameters ${\bf u}_N=(u_1, u_2, \dots , u_N)$ are arbitrary complex numbers, and ${\bf u}^2_N\equiv (u^2_1, u^2_2, \dots , u^2_N)$. The state $|\!\!\Uparrow \rangle$ in (\ref{conwf1}) is the fully polarized one with all spins ``up'': $\mid\Uparrow\rangle
\equiv \bigotimes_{n=0}^M \mid \uparrow
\rangle_n$. The amplitudes in
(\ref{conwf1}) are expressed in terms of the {\it Schur functions}
$S_\bla $ \cite{macd}:
\begin{equation}
S_{\bla} ({\textbf x}_N)\,\equiv\,
\displaystyle{ S_{\bla} (x_1, x_2, \dots , x_N)\,\equiv\, \frac{\det(x_j^{\la_k+N-k})_{1\leq
j, k \leq N}}{\CV({\textbf x}_N)}}\,,
\label{sch}
\end{equation}
in which $\CV ({\textbf x}_N)$  is the Vandermonde determinant
\begin{equation} \CV ({\textbf x}_N)
\equiv
\det(x_j^{N-k})_{1\leq j, k\leq N}
\,=\,
\prod_{1 \leq m<l \leq N}(x_l-x_m)\,.
\label{spxx1}
\end{equation}
The conjugated state-vectors are given by
\begin{equation}\label{conj}
\langle \Psi({\bf v}_N)\mid\,=\,\sum\limits_{\bla \subseteq \{{\CK}^N\}}
\langle \Uparrow \mid \begin{pmatrix}
\prod\limits_{k=1}^N \si_{\mu_k}^{+}\end{pmatrix} S_\bla ({\textbf v}^{-2}_N)\,.
\end{equation}

If parameters $u^2_j\equiv
e^{i\ta_j}$ ($1\le j\le N$) satisfy
the Bethe equations \cite{iz11},
\begin{equation}
e^{i (M+1)\ta_j}=(-1)^{N-1}\,, \quad 1\le j \le N \,,
\label{betheexp}
\end{equation}
then the state-vectors (\ref{conwf1})
become the eigen vectors of the Hamiltonian (\ref {anis1}):
\begin{equation}\label{egv}
{\cal H} \mid\!\Psi (e^{i{\bth }_N/2})\rangle\,=\, E_N({\bth }_N)
\mid\!\Psi (e^{i{\bth }_N/2})\rangle\,.
\end{equation}
Here (and throughout the paper) the notation $\bth_N$ for $N$-tuple $(\ta_{1}, \ta_{2}, \dots , \ta_{N})$ is reserved for the solutions to the Bethe equations (\ref{betheexp}), and $e^{{i\bth}_N} \equiv (e^{\ta_{1}}, e^{\ta_{2}}, \ldots , e^{\ta_{N}})$.
The solutions $\theta_j$ to the Bethe equations (\ref{betheexp}) can be parametrized such that
\begin{equation}
\theta_j = \frac{2\pi
}{M+1}\begin{pmatrix}\displaystyle{
I_j-\frac{N-1}{2}} \end{pmatrix}\,,
\quad 1\le j \le N\,, \label{besol}
\end{equation}
where $I_j$ are integers or half-integers depending on whether $N$ is odd or even. It is sufficient to consider a set of $N$ different numbers $I_j$ satisfying the
condition: $M\geq I_1>I_2> \dots>I_N\geq 0$.

Then the eigen energies in (\ref{egv}) are equal to
\begin{equation}
E_N({\bth}_N)\,=\,- \sum_{j=1}^N\cos\ta_j\,=\,-
\sum_{j=1}^N\cos\begin{pmatrix} \displaystyle{\frac{2\pi }{M+1}
\Bigl(I_j-\frac{N-1}{2}\Bigr)}\end{pmatrix}.
\label{egen}
\end{equation}
The \textit{ground state} of the model is the eigen-state that corresponds to the lowest eigen energy   $E_N(\bth^{\,\rm g}_N)$. It is determined by the solution to the Bethe equations  (\ref{besol}) at $I_j=N-j$:
\begin{equation}\label{grstxx}
  \theta^{\,\rm g}_j \equiv \frac{2\pi
}{M+1}\begin{pmatrix} \displaystyle{\frac{N+1}{2} -j} \end{pmatrix}\,, \quad
1\le j \le N\,,
\end{equation}
and is equal to
\[
E_N(\bth^{\,\rm g}_N)\, = \,-\,\frac{
\,\sin\frac{\pi N}{M+1}}{\sin\frac{\pi}{M+1}}\,.
\]

In the present paper, the two types of the the thermal correlation functions in a system of finite size will be considered. We call them the {\textit {persistence of ferromagnetic string}} and the {\textit {persistence of domain wall}}.
The persistence of ferromagnetic string is related to the projection operator $\bar\varPi_{n}$ that forbids spin ``down'' states on the  first $n$  sites of the chain:
\begin{equation}
\CT ({\bth}^{\,\rm g}_N, n,
\be)\,\equiv\,\frac{\langle \Psi (e^{i{\bth}^{\,\rm g}_N/2}) \!\mid
\bar\varPi_{n}\,e^{-\be {\cal
H}}\,\bar\varPi_{n}
\mid\! \Psi (e^{i{\bth}^{\,\rm g}_N/2}) \rangle
}{\langle \Psi (e^{i{\bth}^{\,\rm g}_N/2}) \!\mid
e^{-\be {\cal H}} \mid\! \Psi (e^{i{\bth}^{\,\rm g}_N/2}) \rangle}\,,\qquad \bar\varPi_{n} \equiv \prod\limits_{j=0}^{n-1}\, \frac{\si^0_j + \si^z_j}{2}\,, \label{ratbe0}
\end{equation}
where ${\cal H}$ and $\bth^{\,\rm g}_N$ are given by (\ref{anis1}) and (\ref{grstxx}), respectively, and $\be\in\BC$.  Some results on this correlation function have been reported in \cite{b5}.

The persistence of domain wall is related to the operator $\bar{\sf F}_{n}$ that creates a sequence of spin ``down'' states on the  first $n$  sites of the chain:
\begin{equation}
{\cal F} (\widetilde{\bth}^{\,\rm g}_{N-n}, n,
\be)\,\equiv\,\frac{\langle
\Psi (e^{i{
\widetilde{\bth}^{\,\rm g}_{N-n}}/2}) \!\mid \bar{\sf
F}^{+}_{n}\,e^{-\be {\cal H}}\,\bar{\sf F}_{n}
\mid\! \Psi (e^{i{
\widetilde{\bth}^{\,\rm g}_{N-n}}/2}) \rangle
}{\langle \Psi (e^{i{
\widetilde{\bth}^{\,\rm g}_{N-n}}/2}) \!\mid e^{-\be {\cal H}} \mid\!
\Psi (e^{i{
\widetilde{\bth}^{\,\rm g}_{N-n}}/2})
\rangle}\,,\qquad \bar{\sf F}_{n}\,
\equiv\,\prod\limits_{j=0}^{n-1} \si^-_j \,,
\label{field0}
\end{equation}
where $\bar{\sf
F}^{+}_{n}$ is the Hermitian conjugated operator acting on the conjugated state-vectors (\ref{conj}), and $\widetilde{\bth}^{\,\rm g}_{N-n}$ is the set of ground state solutions to the Bethe equations (\ref{betheexp})
for the system of $N-n$ particles:
\begin{equation}\label{grstxxn}
   \widetilde \theta^{\,\rm g}_j \equiv \frac{2\pi
}{M+1}\begin{pmatrix} \displaystyle{\frac{N-n+1}{2} -j} \end{pmatrix}\,, \quad
1\le j \le N-n\,.
\end{equation}
We assume that
$\bar\varPi_{0}$ and $\bar{\sf F}_{0}$ are the identity operators so that $\CT
({\bth}^{\,\rm g}_N, 0, \be)=1$ and ${\cal F} (\widetilde{\bth}^{\,\rm g}_{N-n}, 0, \be)=1$.

The calculation of introduced correlation functions will be based on the
Binet--Cauchy formula \cite{gant} adapted for the Schur functions:
\begin{equation}
\CP_{L/n}({\bf y}_N, {\bf
x}_N)\,\equiv\,\sum_{\bla \subseteq
\{(L/n)^N\}} S_{\bla} ({\textbf y}_N) S_{\bla} ({\textbf x}_N)\,=\,
\begin{pmatrix}
\displaystyle{\prod\limits_{l=1}^{N} y_l^n x_l^n }\end{pmatrix}
\frac{\det (T_{k j})_{1\le
k, j\le N}}{\CV ({\textbf y}_N)
\CV ({\textbf x}_N)} \,,
  \label{schr}
\end{equation}
where the summation is over all
partitions $\bla$ satisfying: $L\ge \la_1 \ge \la_2\ge \dots \ge \la_N \ge n$. The Vandermonde determinant (\ref{spxx1}) is used in (\ref{schr}), and the entries $T_{k j}$ are given by:
\begin{equation}
T_{k j}=\frac{1-(x_k
y_j)^{N+L-n}}{1-x_k y_j}\,. \label{hm}
\end{equation}

\section{The correlation functions}
\label{cor:sec2}

In this section we shall calculate the persistence correlation functions of ferromagnetic string and of domain wall. The calculation is natural to start with the derivation of the form-factors of appropriate operators.

\subsection{The Bethe states
and form-factors}

Applying the orthogonality relation
\begin{equation}
\langle \Uparrow \mid \prod\limits_{k=1}^N \si_{\mu_k}^{+}
\prod\limits_{l=1}^N \si_{\nu_l}^{-} \mid \Uparrow \rangle =\prod\limits_{n=1}^N
\dl_{\mu_n \nu_n}\,, \label{tuc3}
\end{equation}
we reduce the calculation of the scalar products of the state-vectors (\ref{conwf1}) and (\ref{conj}) to the  Bine-Cauchy formula (\ref{schr}):
\begin{equation}
\langle \Psi ({\textbf
v}_N)\mid\!\Psi ({\textbf u}_N)\rangle\,=\,
\sum_{\bla \subseteq \{{\CK}^N\}}S_\bla
({\textbf v}^{-2}_N)S_\bla ({\textbf
u}^2_N)\,=\,\frac{\det(T^{\rm o}_{k j})_{1\leq
k, j\leq N}}{{\CV} ({\textbf
u}^2_N) {\CV} ({\textbf v}^{-2}_N)} \,,
\label{spxx}
\end{equation}
where summation is over all partitions $\bla$ with at most $N$ parts, each of which is less or equal to $\CK$.
The entries $T^{\rm o}_{k j}$ in (\ref{spxx}) are  given by (\ref{hm}) taken at $n=0$:
\begin{equation}
T^{\rm o}_{kj} =
\frac{1-(u^2_k/v^2_j)^{M+1}}{1 -
u^2_k/v^2_j}\,. \label{t}
\end{equation}
For $\textbf v_N = \textbf u_N$, this scalar product is equal to the squared ``length'' of the states (\ref {conwf1}): ${\widetilde\CN}^2({\textbf u}_N) \equiv \langle \Psi ({\textbf u}_N) \mid\!\Psi ({\textbf u}_N)\rangle$.

On the solutions  ${\textbf u}^{2}_N={\textbf v}^{2}_N=e^{i\bth_N}$  (\ref{besol})  to the Bethe equations (\ref{betheexp}), the entries (\ref {t}) are equal to
\begin{equation}
T^{\rm o}_{k j}= \frac{\sin\pi
(I_k-I_j)}{\sin\frac{\pi}{M+1}(I_k-I_j)}\,
e^{i\frac{\pi M}{M+1}(I_k-I_j)}\,=\,
(M+1)\dl_{jk}\,, \label{normxx1}
\end{equation}
and for the square of the norm
$\CN^2({\bth}_N)\equiv {\widetilde
\CN}^2(e^{i{\bth}_N/2})$ we have:
\begin{equation}
\CN^2({\bth}_N)\,=\,\displaystyle{
\frac{(M+1)^N}{|\CV (e^{i{\bth}_N})|^2}
=\frac{(M+1)^N}{\prod\limits_{1\leq m<l \leq N} 2 (1-\cos \frac{2
\pi}{M+1}(I_l-I_m))}}\,. \label{normxx}
\end{equation}
Notice, that if $\bth_N$ and ${\bth}^{\prime}_N$ are different sets of the solutions to the Bethe equations, the related eigen vectors are orthogonal \cite{ml4}:
$\langle \Psi (e^{i{{\bth }_N}/2}) \!\mid\! \Psi (e^{i{{\bth }_N^{\prime}}/2}) \rangle =0$. Being a complete and orthogonal set of states, these eigen vectors provide the resolution of the identity operator \cite{KBI1, KBI2}
\begin{equation}
{\BI}\,=\,\sum\limits_{\{{\bth}_N\}}
\CN^{-2}({\bth}_N)
\mid\! \Psi (e^{i{{\bth }_N}/2}) \rangle\,\langle \Psi (e^{i{{\bth }_N}/2}) \!\mid\,, \label{field7}
\end{equation}
where summation is over all different solutions to the Bethe equations (\ref{besol}).

The form-factor of the projector $\bar\varPi_{n}$ (\ref{ratbe0}) is defined by the ratio:
\begin{equation}
\CT ({\textbf v_N}, {\textbf u_N},
n)\,\equiv\,\frac{\langle \Psi ({\textbf
v_N})\mid \bar\varPi_{n}\mid\!\Psi ({\textbf u_N})\rangle }{{\widetilde
\CN}({\textbf v_N}){\widetilde
\CN}({\textbf u_N})}\,. \label{ratio}
\end{equation}
Taking into account (\ref{conwf1}), (\ref{schr}) and (\ref{hm}), we find that the numerator of (\ref{ratio}) is equal to:
\begin{equation}
\langle \Psi ({\textbf v_N})\mid
\bar\varPi_{n}\mid\! \Psi ({\textbf u_N})
\rangle\,=\,\CP_{\CK/n}({\textbf v}^{-2}_N, {\textbf u}^2_N)\,=\,
\begin{pmatrix}
\displaystyle{\prod\limits_{l=1}^{N} \frac{u_l^{2n}}{v_l^{2n}} }\end{pmatrix}
\frac{\det(T_{k j})_{1\leq k, j\leq
N}}{{\CV} ({\textbf u}_N^2){\CV} ({\textbf
v}^{-2}_N)}\,,
\label{spdfpxx0}
\end{equation}
where the entries are equal to
\begin{equation}
T_{kj} =
\frac{1-(u^2_k/v^2_j)^{M+1-n}}{1 -
u^2_k/v^2_j}\,.
\label{spdfpxx}
\end{equation}
Taken at $n=0$, Eq.~(\ref{spdfpxx0})
reproduces the answer for the scalar product (\ref{spxx}).

On the solutions to the Bethe equations (\ref{besol}), the form-factor (\ref{spdfpxx0}) takes the form
\begin{equation}
\langle \Psi (e^{i{{\bth }_N}/2}) \!\mid \bar\varPi_{n}
\mid\! \Psi (e^{i{{\bth }_N}/2}) \rangle\,=\, \frac{(M+1)^N}
{|\CV (e^{{i\bth}_N})|^2} \det\bigl(\delta_{k j}- K_n(\theta_k,
\theta_j)\bigr)_{1\le k, j \le N}\,,
\label{efp1}
\end{equation}
with the entries of matrix $K_n(\theta_k, \theta_j)$ equal to
\begin{equation}
\displaystyle{ K_n(\theta_k,
\theta_j)\,\equiv\,
\frac{e^{i(n-1)(\theta_k-\theta_j)/2}}
{M+1}\,
\frac{\sin\frac{n(\theta_k-\theta_j)}2}
{\sin\frac{\theta_k-\theta_j}2}}\,.
\label{efp2}
\end{equation}
To obtain this answer we have subtracted and added the unity to the numerator of $T_{k j}$ (\ref{spdfpxx}) and used the equality (\ref{normxx1}). Finally, for the form-factor (\ref{ratio}) we obtain
\begin{equation}
\displaystyle{ \CT (e^{i{\bth}_N/2}, e^{i{\bth}_N/2}, n)\,=\, \det\bigl(\dl_{l
m}-K_n(\theta_l, \theta_m)\bigr)_{1\le l, m
\le N}}\,. \nonumber
\end{equation}
In this form it is known as the emptiness formation probability, being the
probability of formation of a string of $n$ consecutive spin ``up'' states \cite{korr, ess, ml5, ml4, iz11}.

Let us consider then the form-factor of the domain wall creation operator $\bar{\sf F}_{n}$ (\ref{field0}):
\begin{equation}
{\cal F} (\textbf v_N, \textbf u_{N-n}, n)\,\equiv\,\frac{\langle \Psi ({\bf v}_N)\mid \bar{\sf
F}_{n} \mid\!\Psi ({\bf u}_{N-n})\rangle }{{\widetilde
\CN}({\textbf v_N}) {\widetilde
\CN}({\textbf
u_{N-n}})}\,. \label{ratiof}
\end{equation}
To calculate this transition element, we first introduce an auxiliary operator
${\sf D}^n ({\bf u})$ acting on an expectation $\langle \cdot \rangle_{\bf u}$ considered as function of ${\bf u}$:
\begin{equation}
{\sf D}^n ({\bf u})\, \langle\cdot\rangle_{\bf u}\,\equiv\,{\sf D}_{u_{N-n+1}, u_{N-n+2}, \dots , u_{N}} \, \begin{pmatrix} \displaystyle{
\frac{{\CV} ({\textbf
u}^2_N)}{{\CV} ({\textbf u}^{2}_{N-n})} \times \langle \cdot \rangle_{\bf u}} \end{pmatrix}\,. \label{dif1}
\end{equation}
Here, ${\sf D}_{u_{N-n+1}, u_{N-n+2}, \dots , u_{N}}$ denotes subsequent $n$-fold application of the differentiation operators ${\sf D}^j_{u_{N-j}}$,
\begin{equation}
{\sf D}_{u_{N-n+1}, u_{N-n+2}, \dots , u_{N}} \,\equiv\,{\sf
D}^{n-1}_{u_{N-n+1}}\,\circ\,{\sf
D}^{n-2}_{u_{N-n+2}}\,\circ\,\dots\, \circ\,{\sf
D}^{0}_{u_{N}}\,,
\label{dif11}
\end{equation}
where
\begin{equation}
{\sf D}^j_{u_{N-j}}\,\equiv\,
\lim_{u^2_{N-j}\to 0}\,\frac{1}{j\,!}\,
\frac{{\d}^j}{{\d}(u^2_{N-j})^j}\,,\qquad 0\le j\le n-1\,.
\label{dif2}
\end{equation}
The definition (\ref{dif1}) implies that
the expectation $\langle\cdot\rangle_{\bf u}$ is first multiplied by the
ratio of the Vandermonde determinants and then differentiated $n$ times.

Now we are ready to formulate the following

\vskip0.3cm \noindent {\bf Proposition~1\,}
\textit{ The action of operator ${\sf
D}^n({\bf u})$ expressed by} (\ref{dif1}),
(\ref{dif11}), (\ref{dif2}) \textit{on the scalar product} $\langle \Psi({\bf
v}_N)\mid\!\Psi ({\bf u}_N)\rangle$
\textit{gives the form-factor of the domain wall creation operator} $\bar{\sf F}_{n}$~(\ref{ratiof}):
\begin{equation}
\langle \Psi ({\bf v}_N)\mid \bar{\sf
F}_{n} \mid\!\Psi ({\bf
u}_{N-n})\rangle\,=\,{\sf D}^n({\bf
u}) \langle
\Psi ({\bf v}_N)\mid \Psi ({\bf
u}_N)\rangle\,.
\label{dif3}
\end{equation}

\vskip0.3cm \noindent {\bf Proof\,} From the definition of the state-vectors
(\ref{conwf1}) and of the operator $\bar{\sf F}_{n}$ (\ref{field0}) we obtain the representation of the form-factor:
\begin{equation}
\langle \Psi ({\bf v}_N)\mid \bar{\sf
F}_{n} \mid\!\Psi ({\bf u}_{N-n})\rangle\,=\,\begin{pmatrix}
\displaystyle{\prod\limits_{l=1}^{N-n} u^{2n}_l }\end{pmatrix} \sum_{\bla
\subseteq \{\CK^{N-n}\}} S_{\hat\bla}
({\textbf v}^{-2}_N) S_{\bla} ({\textbf
u}^2_{N-n})\,, \label{field41}
\end{equation}
where summation runs over the partitions
$\bla$ of the length $N-n$: $\CK \ge\la_1 \ge\la_2 \ge\,\dots\, \ge \la_{N-n}\ge 0$. The parts of the partition $\hat\bla$ are $\hat\la_p=\la_p$ at $1 \le p \le N-n$, and $\hat\la_{N-n+1} = \hat\la_{N-n+2} = \dots = \hat\la_{N}=0$. The corresponding strict
partitions are given by $\hat\bmu =\hat\bla +\bdl_N$ and $\bmu =\bla +\bdl_{N-n}$.

To derive (\ref{field41}), let us act by $\bar{\sf F}_{n}$ on the state $\mid\!\Psi ({\bf u}_{N-n})\rangle$ given by (\ref{conwf1}) with the summation taken over $\bla \subseteq \{(\CK+n)^{N-n}\}$. The operator $\bar{\sf F}_{n}$ acts non-trivially only on those vectors in $| \Psi({\bf
u}_{N-n})\rangle$, that do not contain spin ``down'' states on the first $n$ sites:
\begin{multline}\label{fs}
\bar{\sf F}_{n} \mid\!\Psi ({\textbf u}_{N-n})\rangle =
\sum\limits_{{\bla} \subseteq
\{((\CK+n)/n)^{N-n}\}}
S_\bla ({\textbf u}^{2}_{N-n})
\begin{pmatrix}
\prod\limits_{l=0}^{n-1} \si_{l}^{-}\end{pmatrix} \begin{pmatrix}
\prod\limits_{k=1}^{N-n} \si_{\mu_k}^{-}\end{pmatrix} \mid
\Uparrow \rangle \\
=\,
\begin{pmatrix}
\displaystyle{\prod\limits_{l=1}^{N-n} u^{2n}_l }\end{pmatrix} \sum_{\bla
\subseteq \{\CK^{N-n}\}}
S_\bla ({\textbf u}^{2}_{N-n}) \begin{pmatrix}
\prod\limits_{l=0}^{n-1} \si_{l}^{-}\end{pmatrix} \begin{pmatrix}
\prod\limits_{k=1}^{N-n} \si_{\mu_k}^{-}\end{pmatrix} \mid
\Uparrow \rangle\,.
\end{multline}
We have used the definition of the Schur function (\ref{sch}) to obtain the last equality.
Applying the orthogonality relation (\ref{tuc3}) to right-hand side of (\ref{fs}), we find that right-hand side of (\ref{field41}) indeed holds true.

Eventually, direct evaluation of
right-hand side of (\ref{dif3}) leads to right-hand side of
(\ref{field41}) provided that the scalar product is represented through the Schur functions according to (\ref{spxx}). $\Box$

\vskip0.3cm Proposition~1 enables us to obtain two summation rules for the products of the Schur functions, which are crucial in establishing of the combinatorial results for the correlation functions in question. So, we formulate the following
\vskip0.3cm \noindent {\bf Proposition 2\,}
\textit{The following sums of products of the Schur functions take place}:
\begin{eqnarray}
&&\sum\limits_{\bla \subseteq \{\CK^{N-n}\}}
S_{\hat\bla} ({\textbf v}^{-2}_N) S_{\bla} ({\textbf u}^2_{N-n})\,=\,\displaystyle{
\begin{pmatrix} \displaystyle{\prod\limits_{l=1}^{N-n}
u^{-2n}_l } \end{pmatrix} \frac{\det(\bar T_{kj})_{1\le k, j\le
N}}{{\CV} ({\textbf u}^2_{N-n}){\CV} ({\textbf v}^{-2}_N)}}\,,
\label{field43}\\[0.3cm]
&&\sum\limits_{\bla \subseteq \{\CK^{N-n}\}}S_{\bla}
({\textbf v}^{-2}_{N-n}) S_{\hat\bla} ({\textbf u}^2_N)\,=\,\begin{pmatrix} \displaystyle{\prod\limits_{l=1}^{N-n}
v^{2n}_l}\end{pmatrix}
 \displaystyle{ \frac{\det(\tilde
T_{kj})_{1\le k, j\le
N}}{{\CV} ({\textbf v}^{-2}_{N-n}){\CV} ({\textbf u}^2_N)}}
\label{field51}\,,
\end{eqnarray}
\textit{where the entries of the matrices} $(\bar T_{kj})_{1\le k, j\le
N}$ \textit{and} $(\tilde T_{kj})_{1\le k, j\le
N}$ \textit{are}:
\begin{equation}
\displaystyle{ \begin{array}{ll} \bar
T_{kj} = T^{\rm o}_{kj}\,, & 1\le k \le
N-n,\,\quad\qquad 1\le j \le N\,,\\[0.2cm] \bar T_{kj} = v_j^{-2(N-k)}\,, & N-n+1\le k \le
N,\,\quad 1\le j \le N\,,
\end{array}} \label{field42}
\end{equation}
\textit{and}
\begin{equation}
\displaystyle{ \begin{array}{ll} \tilde
T_{kj} = T^{\rm o}_{kj}\,, & 1\le k \le
N,\,\qquad 1\le j \le
N-n\,,\\[0.2cm] \tilde T_{kj} = u_j^{2(N-k)}\,, & 1\le k \le N,\,\qquad\,N-n+1\le j \le N\,.
\end{array}}
\nonumber
\end{equation}

\vskip0.3cm \noindent {\bf Proof\,}
Let us calculate right-hand side of (\ref{dif3}) using the determinantal form of the scalar product
(\ref{spxx}):
\begin{equation}
\langle \Psi ({\bf v}_N)\mid \bar{\sf
F}_{n} \mid\!\Psi ({\bf
u}_{N-n})\rangle\,=\,
{\sf D}^n({\bf
u})\,\begin{pmatrix} \displaystyle{
\frac{\det(T^{\rm o}_{k
j})_{1\leq k, j\leq N}}{{\CV} ({\textbf
u}^2_N){\CV} ({\textbf v}^{-2}_N)} }\end{pmatrix} \,.
\label{field3}
\end{equation}
Taking into account the explicit form of the entries $T^{\rm o}_{k j}$ (\ref{t}) we obtain:
\begin{equation}
\langle \Psi ({\bf v}_N)\mid \bar{\sf
F}_{n} \mid\!\Psi ({\bf
u}_{N-n})\rangle\,=\,\displaystyle{ \frac{\det(\bar T_{kj})_{1\le k, j\le
N}}{{\CV} ({\textbf u}^2_{N-n}) {\CV} ({\textbf v}^{-2}_N)}}\,,
\label{field4}
\end{equation}
where the matrix $\bar T$ is given by  (\ref{field42}). Since right-hand sides of (\ref{field41}) and (\ref{field4}) mutually coincide, the relation on the Schur functions (\ref{field43}) (which is of the type of (\ref{schr})) holds true.

Repeating the arguments of Proposition~1, we find that the form-factor of the conjugated operator $ \bar{\sf
F}_{n}^{+}$ is equal to
\begin{equation}\label{ffcf}
\langle \Psi ({\bf v}_{N-n})\mid \bar{\sf
F}_{n}^{+} \mid\!\Psi ({\bf
u}_N)\rangle\,=\,
\begin{pmatrix} \displaystyle{\prod\limits_{l=1}^{N-n}
v^{-2n}_l}\end{pmatrix}
 \displaystyle
\sum_{\bla \subseteq \{\CK^{N-n}\}}S_{\bla}
({\textbf v}^{-2}_{N-n}) S_{\hat\bla} ({\textbf
u}^2_N),
\end{equation}
and to
\begin{equation}
\langle \Psi ({\bf v}_{N-n})\mid \bar{\sf F}_{n}^{+} \mid\!\Psi ({\bf
u}_N)\rangle\,=\,{\sf D}^n({\bf
v}^{-1})\langle \Psi ({\bf v}_N)\mid \Psi ({\bf u}_N)\rangle\,.
\label{field5}
\end{equation}
Respectively, we have for the Schur functions the equality Eq.~(\ref{field51}) as well.
However, the Eq.~(\ref{field51}) can directly  be obtained from Eq.~(\ref{field43}) by changing notations. $\Box$

\subsection{Persistence of
ferromagnetic string} \label{cor:sec32}

Let us recall the main relations concerning the persistence of ferromagnetic string and its relationship with the problem of
\textit{vicious walkers} in the
\textit{random turns model} \cite{1}. The problem of enumeration of the vicious walkers is actively investigated \cite{2, 3, 4, 6, 7, 8, 9, 13}. The random walks across one-dimensional periodic lattice are closely related to the correlation functions of the $XX0$ magnet \cite{b1, b2, b4}.

Taking into account the explicit form of the state vectors (\ref{conwf1}) and (\ref{conj}), we obtain the following answer \cite{b5} for the matrix element
\begin{equation}
\begin{array}{rcl}
&&\langle \Psi ({\textbf v}_N) \mid
\bar\varPi_{n}\, e^{-\be {\cal H}}\,\bar\varPi_{n}\mid\!\Psi ({\textbf u}_N)\rangle \\[0.5cm]
&&=\sum \limits_{{\bla^L},\,{\bla^R}
\subseteq \{(\CK/n)^N\}}
S_{{\bla^L}}({\textbf v}^{-2}_N)
S_{{\bla^R}}({\textbf u}^2_N)
F_{{\bmu^L};\,{\bmu^R}} (\be)\,,
\label{ratbe1}
\end{array}
\end{equation}
parametrized by arbitrary ${\textbf u}_N$ and ${\textbf v}_N$. The range of two independent summations over ${\bla}^L$ and ${\bla}^R$ is defined in (\ref{schr}), and ${\bmu}^{L, R}={\bla}^{L, R}+{\bdl}_N$ denote the
corresponding strict partitions. The transition amplitude:
\begin{equation}
F_{{\bmu^L};\,{\bmu^R}}(\be)\,\equiv\,
\langle\Uparrow \mid
\begin{pmatrix} \displaystyle{
\prod\limits_{l=1}^{N}
\si_{\mu^L_l}^{+}}
\end{pmatrix} e^{-\be {\cal H}} \begin{pmatrix} \displaystyle{
\prod\limits_{p=1}^{N} \si_{\mu^R_p}^{-}}\end{pmatrix} \mid
\Uparrow\rangle\,, \label{ratbe2}
\end{equation}
is related to enumeration of paths of $N$ vicious walkers moving
across the sites of a one-dimensional chain \cite{b1, b2, b4, b5}.
The expression (\ref{ratbe1}) at $\be=0$ is in agreement with (\ref{spdfpxx0}).

The transition amplitude (\ref{ratbe2})
satisfies the difference-differential
equation derived in~\cite{b2}. The corresponding solution has the determinantal form:
\begin{equation}
F_{\mu^L_1,\,\mu^L_2,\,\dots,\,\mu^L_N;\,
\mu^R_1,\,\mu^R_2,\,\dots,\,\mu^R_N} ({\be})\,=\, \det \bigl(F_{\mu^L_k;\,\mu^R_l}({\be}) \bigr)_{1\leq k, l\leq N}\,,
\label{ratbe4}
\end{equation}
where the entries
\begin{equation}
F_{k;\,l}(\be)\,\equiv\,\displaystyle{
\frac{1}{M+1} \sum\limits_{s=0}^M
e^{\be\cos\phi_s}\,e^{i \phi_s(k-l)}}\,, \qquad \phi_s=\frac{2\pi}{M+1}
\Bigl(\displaystyle{
s-\frac{M}{2} }\Bigr)\,,
\label{ratbe6}
\end{equation}
are the transition amplitudes $F_{k;\,l}({\be})=\langle \Uparrow|\si_{k}^{+} e^{-\be {\cal H}} \si_{l}^{-}|\Uparrow \rangle$, which are (\ref{ratbe2}) for $N=1$. They may be considered as generating functions of single pedestrian traveling between l$^{\rm th}$ and k$^{\rm th}$ sites of periodic chain.

The substitution of (\ref{ratbe6}) into (\ref{ratbe4}) allows us to express the transition amplitude (\ref{ratbe2}) through the Schur functions (\ref{sch}) and the Vandermonde determinants (\ref{spxx1}) \cite{b1}:
\begin{equation}
F_{{\bmu^L};\,{\bmu^R}}(\be)\,=\,
\displaystyle{\frac{1}{(M+1)^N}
\sum\limits_{\{{\bphi}_N\}}
e^{-\be E_N({\bphi}_N)}}|{\CV} (e^{i {\bphi}_N})|^2\,
S_{{\bla^L}}(e^{i {\bphi}_N})
S_{{\bla^R}}(e^{-i {\bphi}_N})\,,
\label{ratbe7}
\end{equation}
where the summation is over $N$-tuples ${\bphi}_N \equiv (\phi_{k_1}, \phi_{k_2}, \dots , \phi_{k_N})$ labeled by the integers $k_i$, $1\le i\le N$, respecting $M\ge k_1 > k_2 \dots > k_N\ge 0$. The energy $E_N({\bphi}_N)$ is defined by (\ref{egen}). Substituting
(\ref{ratbe7}) into (\ref{ratbe1}) and applying the Binet--Cauchy formula (\ref{schr}), we obtain \cite{b5}:
\begin{eqnarray}
&&\langle \Psi ({\textbf v}_N)\mid \bar
\varPi_{n}\, e^{-\be {\cal
H}}\,\bar\varPi_{n}\mid\!\Psi ({\textbf u}_N) \rangle\,=
\nonumber\\[0.3cm]
&&=\,\displaystyle{\frac{1}{(M+1)^N}
\sum\limits_{\{{\bphi}_N\}} e^{
-\be E_N({\bphi}_N)}}
|{\CV} (e^{i {\bphi}_N})|^2\,
{\CP}_{\CK/n}({\textbf v}^{-2}_N,
e^{i {\bphi}_N})\,{\CP}_{\CK/n}(e^{-i {\bphi}_N},
{\textbf u}^{2}_N) \label{ratbe80} \\[0.0cm]
&&=\,\displaystyle{
\frac{1}{{\CV} ({\textbf
u}^2_N){\CV} ({\textbf v}^{-2}_N)} \det
\begin{pmatrix} \displaystyle{ \sum\limits_{k, l=n}^{M}
F_{k;\,l}(\be)\,\frac{u_i^{2
l}}{v_j^{2k}} } \end{pmatrix}_{1\le i,j \le N}}\,,
\label{ratbe8}
\end{eqnarray}
where $\CP_{L/n}({\bf y}_N, {\bf x}_N)$ and $F_{k;\,l}(\be)$ are defined by (\ref{schr}) and (\ref{ratbe6}), respectively. At $\be=0$, the expression (\ref{ratbe8}) transfers into
(\ref{spdfpxx0}). Notice, that for $n=0$ the operator $\bar\varPi_{n}=1$ and Eq.~(\ref{ratbe8}) gives the answer for the matrix element $\langle \Psi ({\textbf v}_N)\mid\, e^{-\be {\cal H}}\,\mid\!\Psi ({\textbf u}_N) \rangle $.

Taking into account that
\begin{equation}
\langle \Psi ({\bf v}_N)\mid e^{-\be {\cal H}}\mid\! \Psi (e^{i{{\bth }_N}/2}) \rangle\,=\, \langle
\Psi ({\bf v}_N) \mid\! \Psi (e^{i{{\bth }_N}/2}) \rangle\, e^{-\be E_N ({\bth}_N) }\,, \label{ratbe91}
\end{equation}
where the eigen energy $E_N ({\bth}_N)$ is given by (\ref{egen}), we obtain from
(\ref{ratbe8}) the answer for the
persistence of ferromagnetic string
(\ref{ratbe0}),
\begin{equation}
\CT ({\bth}^{\,\rm g}_N, n,
\be)\,=\,\displaystyle{ \frac{e^{\be E_N
({\bth}^{\,\rm g}_N)}}{(M+1)^{N}}\,\det
\begin{pmatrix} \displaystyle{ \sum\limits_{k, l=n}^{M}
F_{k;\,l}(\be)\,e^{i(l\ta^{\,\rm
g}_i - k \ta^{\,\rm g}_j)}}\end{pmatrix}_{1\le i, j \le
N}}\,, \label{ratbe92}
\end{equation}
with the ground-state energy $E_N
({\bth}^{\,\rm g}_N)$ given by (\ref{grstxx}).

From the relation (\ref{ratbe80})  follows the representation of the correlation function that we will use in the analysis of its asymptotical behaviour:
\begin{eqnarray} \label{ratbe131}
\CT ({\bth}^{\,\rm g}_N, n,
\be)\,=\,\frac{1}{{\CN}^2({\bth}^{\,\rm g}_N) (M+1)^{N}}\,
\displaystyle{\sum\limits_{\{{\bth}_N\}} e^{-\be (E_N({\bth}_N)-E_N(\bth^{\,\rm g}_N))}} \nonumber \\
\times\,\bigl|\displaystyle{
{\CV}(e^{i{\bth}_N})\,
{\CP}_{\CK/n}(e^{-i{\bth}_N}, e^{i {\bth}^{\,\rm
g}_N})\bigr|^2\,,}
\end{eqnarray}
where ${\CN}({\bth}^{\,\rm g}_N)$ is the norm (\ref{normxx}) of the ground state defined by (\ref{grstxx}), and
${\CP}_{\CK/n} (e^{-i{\bth}_N}, e^{i
{\bth}^{\,\rm g}_N})$ is  (\ref{schr}) on the solutions to Bethe equations (\ref{betheexp}).

The approach of the calculation of the
persistence of ferromagnetic string used in this section allowed us to demonstrate the combinatorial nature of this correlation function. Naturally, the same answers could be obtained by insertion of the identity operator into the left-hand side of (\ref{ratbe1}). In this way we shall calculate the persistence of domain wall in the next section.

\subsection{Persistence of
domain wall}

To calculate the persistence of domain wall we insert the resolution of unity operator (\ref{field7}) into the numerator of (\ref{field0}) taken at arbitrary parametrization:
\begin{eqnarray}
&&\langle \Psi ({\bf v}_{N-n})\mid \bar{\sf
F}_{n}^{+}\,e^{-\be {\cal H}}\, \bar{\sf F}_{n}
\mid\!\Psi ({\bf u}_{N-n})
\rangle\,= \label{field6}\\[0.3cm]
&&=\,\sum\limits_{\{{\bth}_N\}} \langle
\Psi ({\bf v}_{N-n})\mid \bar{\sf F}_{n}^{+}
\mid\! \Psi (e^{i{{\bth }_N}/2}) \rangle\, \langle
\Psi (e^{i{{\bth }_N}/2}) \!\mid \bar{\sf F}_{n}
\mid\!\Psi ({\bf
u}_{N-n})\rangle\,\frac{e^{-\be E_N
({\bth}_N)}}{\CN^{2}({\bth}_N)} \label{field81}
\\[0.3cm]
&&=\,{\sf D}^n({\bf u})\,{\sf D}^n({\bf
v}^{-1}) \langle \Psi ({\textbf v}_N)\mid e^{-\be {\cal H}}\mid\!\Psi ({\textbf
u}_N)\rangle \,. \label{field8}
\end{eqnarray}
The decomposition (\ref{field81}) transfers into (\ref{field8}) provided that (\ref{dif3}) and (\ref{field5}) are used for each of form-factors in (\ref{field81}).

The substitution of the equality (\ref{ratbe8}), taken at $n=0$, into (\ref{field8}) gives
\begin{equation}
\begin{array}{rcl}
&&\langle \Psi ({\bf v}_{N-n})\mid \bar{\sf
F}_{n}^{+}\,e^{-\be {\cal H}}\, \bar{\sf F}_{n} \mid\!\Psi ({\bf u}_{N-n})\rangle \,=\,
\displaystyle{\frac{1}{{\CV} ({\textbf
u}^2_{N-n}){\CV} ({\textbf
v}^{-2}_{N-n})}}\\[0.6cm]
&&\times\,{\sf D}_{v^{-1}_{N-n+1}, v^{-1}_{N-n+2}, \dots , v^{-1}_{N}} \circ
{\sf D}_{u_{N-n+1}, u_{N-n+2}, \dots , u_{N}} \det \begin{pmatrix} \displaystyle{
\sum\limits_{k, l=0}^{M}
F_{k;\,l}(\be)\,
\frac{u_i^{2l}}{v_j^{2k}}} \end{pmatrix}_{1\le i,j \le N} \,,
\end{array}
\label{field12}
\end{equation}
where ${\sf D}_{u_{N-n+1}, u_{N-n+2}, \dots , u_{N}}$ is given by (\ref{dif11}); ${\sf D}_{v^{-1}_{N-n+1}, v^{-1}_{N-n+2}, \dots , v^{-1}_{N}}$ is defined analogously. After the
differentiations the representation
(\ref{field12}) takes the form:
\begin{equation}
\langle \Psi ({\bf v}_{N-n})\mid \bar{\sf
F}_{n}^{+}\,e^{-\be {\cal H}}\, \bar{\sf F}_{n}
\mid\!\Psi ({\bf u}_{N-n})\rangle
\,=\,\displaystyle{\frac{1}{
{\CV} ({\textbf u}^2_{N-n}){\CV} ({\textbf v}^{-2}_{N-n})}}\, {\det}\displaystyle{
\begin{pmatrix} {\bf A}\, {\bf B}\\
{\bf C}\, {\bf D}\end{pmatrix}} \,,
\nonumber
\end{equation}
where ${\bf A}$, ${\bf B}$, ${\bf C}$, ${\bf D}$ are the matrices with the entries:
\begin{eqnarray}
& A_{i j}\equiv\sum\limits_{k, l=0}^{M}
F_{k;\,l}(\be)\,\displaystyle{ \frac{u_i^{2l}}{v_j^{2k}}}\,,
& 1\le i,j \le
N-n\,,
\nonumber \\[0.2cm]
& B_{i j}\equiv\sum\limits_{l=0}^{M}
F_{n-j;\,l}(\be)\,u_i^{2l}\,,
& 1\le i\le N-n\,,\,\, 1\le j \le n \,,
\nonumber\\[0.2cm]
&
C_{i j}\equiv\sum\limits_{l=0}^{M}
F_{l;\, n-i}(\be)\,v_j^{-2l}\,,
& 1\le i\le n\,,\,\, 1\le j \le N-n \,,
\nonumber\\[0.2cm]
& D_{i j}\equiv F_{n-i;\,n-j}(\be)\,,
& 1\le i, j\le n\,. \nonumber
\end{eqnarray}
Finally, we obtain the answer for the persistence of domain wall (\ref{field0}):
\begin{equation}
{\cal F}(\widetilde{\bth}_{N-n}^{\,\rm g}, n, \be)\,= \,
\frac{e^{\be E_{N-n} (\widetilde{\bth}^{\,\rm
g}_{N-n})}}{(M+1)^{N-n}}\,{\det} \displaystyle{
\begin{pmatrix} {\bf
A}|_{\widetilde{\bth}^{\,\rm g}}\, {\bf
B}|_{\widetilde{\bth}^{\,\rm g}}\\ {\bf
C}|_{\widetilde{\bth}^{\,\rm g}}\, {\bf
D}|_{\widetilde{\bth}^{\,\rm g}}\end{pmatrix}} \,,
\nonumber
\end{equation}
where ${\bf A}|_{\widetilde{\bth}^{\,\rm g}} \equiv
\lim_{{\bf u}^2, {\bf
v}^2 \rightarrow \exp (i\widetilde {\bth}_{N-n}^{\,\rm g})} {\bf A}$ (the same for ${\bf B}$, ${\bf C}$, and ${\bf D}$).

The explicit expression for the form-factor (\ref{field41}) allows us to express the persistence of the domain wall in terms of Schur functions starting with the relation  (\ref{field81}):
\begin{eqnarray}
{\cal F}(\widetilde{\bth}_{N-n}^{\,\rm g}, n, \be)\,=\,\displaystyle{\frac{1}{{\CN}^2
(\widetilde{\bth}_{N-n}^{\,\rm g}) (M+1)^{N-n}}
\sum\limits_{\{\widetilde{\bth}_{N-n}\}} e^{-\be
(E_{N-n}(\widetilde{\bth}_{N-n})- E_{N-n}(\widetilde{\bth}_{N-n}^{\,\rm g}))}}
\nonumber\\[0.3cm]
 \times\, \Bigr|{\CV} (e^{i \widetilde{\bth}_{N-n}})
\sum_{{\bla} \subseteq \{\CK^{N-n}\}}
S_{{\hat\bla}} (e^{-i \widetilde{\bth}_{N-n}}) S_{\bla} (e^{i
\widetilde{\bth}_{N-n}^{\,\rm g}})\Bigl|^2\,.
\label{field181}
\end{eqnarray}
where the summation is over all solutions to the Bethe equation (\ref{betheexp}), and  $\widetilde{\bth}^{\,\rm g}_{N-n}$ is the ground state solution of the system of $N-n$ particles (\ref{grstxxn}).

\section{$q$-Binomial determinants and boxed plane partitions}
\label{cor:sec3}

Boxed plane partitions and $q$-binomial determinants are the important notions that will allow us to give the combinatorial interpretation of the asymptotical behavior of the correlation functions. Proposition~3 formulated in this section provides the determinantal formulas which enable the connection between the form-factors and enumeration of boxed plane partitions as well as of certain non-intersecting lattice paths.

\subsection{$q$-Binomial determinants}
\label{cor:sec42}

The scalar product of the state-vectors (\ref{spxx}), as well as the form-factors  (\ref{spdfpxx0}) and (\ref{field4}), are connected with the generating functions of boxed plane partitions ({\rm AI}.1)
and ({\rm AI}.3) (see Appendix~I). This connection takes place under special $q$-parametrization of the free variables ${\bf u}_N$ and ${\bf v}_N$, and appropriate formulas are given by the statements of Proposition~3.
Before turning to Proposition~3, we shall remind essential notions concerning the \textit{$q$-binomial determinants}, \cite{car}.

To study the asymptotical behavior of the introduced correlation functions, we need to calculate the determinant of the matrix $(\bar {\sf
T})_{1\le j, k \le N}$ defined by Eqs.~(\ref{field42}) though taken under the special \textit{q}-parametrization,
\begin{equation}
{\bf v}_N^{-2}={\bf q}_N\equiv (q, q^2, \dots,
q^N)\,,\qquad {\bf u}_N^{2}={\bf q}_N/q = (1,
q, \dots, q^{N-1})\,. \label{rep21}
\end{equation}
For the arbitrary $P$ and $L\leq N$, these entries will take the form:
\begin{equation}
\displaystyle{ \begin{array}{ll} \bar {\sf
T}_{kj} = \displaystyle{
\frac{1-q^{(P+1)(j+k-1)}}{1-q^{j+k-1}}}\,,
& 1\le k \le L,\,\qquad\quad 1\le j \le N\,,\\
[0.4cm] \bar {\sf T}_{kj} = q^{j(N-k)}\,, &
L+1 \le k \le N,\,\quad 1\le j \le N\,.
\end{array}} \label{rep2}
\end{equation}
This square matrix $(\bar {\sf
T})_{1\le j, k \le N}$ consists of two
blocks of the sizes $L\times N$ and
$(N-L)\times N$. When $L=N$, it consists of one block and is the matrix (\ref{t}) under the \textit{q}-parametrization. It seems appropriate to call the determinant, $\det\bar {\sf T}$, as the
Kuperberg-type determinant (see \cite{kup}, where the problem of enumeration of alternating sign matrices has been investigated).

The $q$-binomial determinant
$\begin{pmatrix}
{\bf{ a}}\\
{\bf{ b}}
\end{pmatrix}_q$ is defined by
\begin{equation}
\begin{pmatrix}
{\bf{ a}}\\
{\bf{ b}}
\end{pmatrix}_q\, \equiv \begin{pmatrix}
a_1, & a_2, & \dots & a_S\\
b_1, & b_2, & \dots & b_S
\end{pmatrix}_q\,\equiv\, \displaystyle{
 \det \begin{pmatrix}
\begin{bmatrix}
a_j\\
b_i
\end{bmatrix}
\end{pmatrix}_{1\le i, j \le S}}
\label{bd0}\,,
\end{equation}
where ${\bf{a}}$ and ${\bf{b}}$ are ordered tuples: $0\le a_1< a_2<\cdots<
a_S$ and $0\le b_1< b_2<\cdots< b_S$.
The entries $\begin{bmatrix}
a_j\\
b_i
\end{bmatrix}$ are the \textit{$q$-binomial
coefficients} (see Appendix~II). In the limit $q\to 1$, the $q$-binomial coefficients are replaced by
the binomial coefficients
$\begin{pmatrix} a_j \\ b_i\end{pmatrix}$.
Then, the $q$-binomial determinant
(\ref{bd0}) is transformed to the
\textit{binomial determinant}:
\begin{equation}
\begin{pmatrix}
{\bf{ a}}\\
{\bf{ b}}
\end{pmatrix}\,\equiv \begin{pmatrix}
a_1, & a_2, & \dots & a_S\\
b_1, & b_2, & \dots & b_S
\end{pmatrix}\,=\, \displaystyle{
 \det \begin{pmatrix}
\begin{pmatrix}
a_j\\
b_i
\end{pmatrix}
\end{pmatrix}_{1\le i, j \le S}}\,.
 \label{bd02}
\end{equation}
The binomial determinant (\ref{bd02}) is
non-negative and is positive provided $b_i\le a_i$, $\forall i$, \cite{ges}.

Now we are ready to formulate the following

\vskip0.3cm \noindent {\bf Proposition 3\,}
\textit{Let the square matrix $(\bar {\sf T})_{1\le j,k \le N}$, consisting of two
blocks of the sizes $L\times N$ and
$(N-L)\times N$, be defined by the entries} (\ref{rep2}) \textit{with $\frac P2 < N< P$. Then, the determinant of} $(\bar {\sf T})_{1\le j,k \le N}$ \textit{is given by either of the following relations}:
\begin{align}
& q^{-\frac L2(L-1)(N-L)}\,\frac{\det(\bar
{\sf T})_{1\le j,k \le N}}{{\CV}({\bf
q}_{N}) {\CV}({\bf q}_{L}/q)} \nonumber \\[0.3cm]
& =\,q^{-\frac
N2({\CP}-1) {\CP}}
\begin{pmatrix}
L+N, & L+N+1, & \dots & L+N+{\CP}-1\\
L, & L+1, & \dots & L+{\CP}-1
\end{pmatrix}_q \label{rep30}\\[0.3cm]
& =\,\displaystyle{\prod_{k=1}^{\CP}
\prod_{j=1}^{L}
\frac{1-q^{j+k+N-1}}{1-q^{j+k-1}}}\,=\,Z_q
(L, N, {\CP})\,,
\label{rep31}
\end{align}
\textit{where $\CP\equiv P-N+1$},
\textit{and  $Z_q (L, N, {\CP})$ is the generating function of plane partitions} ({\rm AI}.1) \textit{contained in a box} ${\cal B}(L, N, {\CP})$.

\vskip0.3cm\noindent {\bf Proof\,} Appendix~II contains the proof of (\ref{rep30}) and (\ref{rep31}). The proof is based on the theory of the symmetric functions. The statements of Proposition~3 are valid for $1\le L\le N$. However, a formal relation can be written for $L=0$ also:
\[\det\bar {\sf T} =q^{-\frac N2({\CP}-1)
{\CP}}{\CV}({\bf q}_{N}) \begin{pmatrix}
N, & N+1, & \dots & N+{\CP}-1\\
0, & 1, & \dots & {\CP}-1
\end{pmatrix}_q\,.\]
In this case, the \textit{q}-binomial determinant is equal to $q^{{\frac
N2({\CP}-1){\CP}}}$ (its
evaluation is in (AII.17)), and $\det\bar {\sf T}$ is nothing but the Vandermonde determinant.  $\Box$
\vskip0.3cm\noindent {\sf Comment \,}
In the limit when $q\rightarrow 1$, the $q$-binomial determinant (\ref{rep30}) transfers into binomial determinant while the generating function (\ref{rep31}) into the number of plane partitions $A(L, N, {\CP})$ ({\rm AI}.2) in a box ${\cal B}(L, N, {\CP})$. Thus, we have:
\begin{equation}\label{binplp}
  \begin{pmatrix}
L+N, & L+N+1, & \dots & L+N+{\CP}-1\\
L, & L+1, & \dots & L+{\CP}-1
\end{pmatrix}=A(L, N, {\CP})\,.
\end{equation}

The number of ways to travel from $(0, 0)$ to $(n, m)$ on a square lattice making elementary steps only to the north and to the east is equal to the binomial
coefficient $\begin{pmatrix} n+m \\ m \end{pmatrix}$. These ways are called the \textit{lattice paths}.
It was found in \cite{ges} that the binomial determinant (\ref{bd02}) is equal to the number of self-avoiding lattice paths $w_1, w_2, \ldots , w_S$ on a square lattice such that $w_i$ goes from $A_i=(0, a_i)$ to $B_i=(b_i, b_i)$, $1\leq i \leq S$. In the considered case, Eq.~(\ref{binplp}), the binomial determinant is equal to the number of self-avoiding lattice paths starting at $A_i = (0, N+L+i-1)$ and terminating at $B_i = (L+i-1, L+i-1)$, $1\leq i \leq {\CP}$. Because of the boundary conditions, this number of self-avoiding paths is equal to the number of self-avoiding paths starting at $C_i=(i, N+L+i-1)$ and terminating at $B_i$. The latter configurations are known as \textit{watermelons} \cite{1}. There exists bijection between watermelons and plane partitions confined in a box of finite size \cite{4}, and it provides the combinatorial proof of (\ref{binplp}) (see Figure~2).
\begin{figure} [h]
\center
\includegraphics {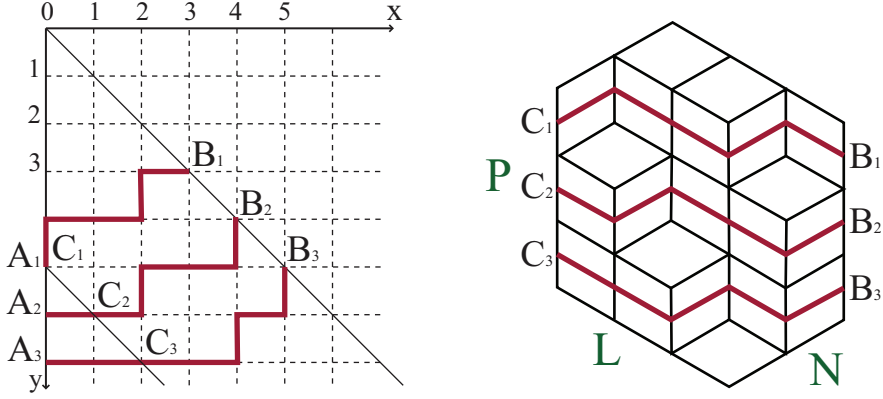}
\caption{Self-avoiding lattice paths, watermelon configuration and plane partition with gradient lines. }
\end{figure}

Described lattice paths from $(0, 0)$ to $(n, m)$ contain in the rectangle $n\times m$. If we place unit squares above and to the left of these lattice paths then the number of lattice paths is equal to the number of different ways to fit the Young diagrams with the largest part at most $n$ and with at most $m$ parts into an $n\times m$ rectangle. The $q$-binomial coefficient $\begin{bmatrix} n+m \\ m \end{bmatrix}$ ({\rm AII}.3) is the generating function of these Young diagrams (lattice paths), and each diagram comes with the weight $q^{|\bla |}$ \cite{stanl}. The weight of the lattice paths that terminate at $B_i=(L+i-1, L+i-1)$ and start at $A_i=(0, N+L+i-1)$ or $C_i=(i, N+L+i-1)$, respectively, differs on common factor $q^{\frac N2 ({\CP}-1) {\CP}}$. Hence, the $q$-binomial determinant (\ref{rep30}) is equal to the generating function of the plane partitions ({\rm AI}.1) contained in ${\cal B}(L, N, {\CP})$ multiplied on the common factor. The algebraic proof of this statement is given in Appendix~II. $\Box$

\subsection{The form-factors and enumeration of boxed plane partitions}
\label{cor:sec4}

Consider now the scalar product $\langle \Psi({\bf v}_{N})\,|\,\Psi({\bf u}_{N})\rangle$ (\ref{spxx}) under the
$q$-parametrization (\ref{rep21}). The entries of the matrix $T^{\rm o}$ (\ref{t}) are (\ref{rep2}) taken for $L=N$, $P=M$:
\begin{equation}
\begin{array}{l}
\langle \Psi ({\bf q}^{- \frac
12}_N)\mid\!\Psi (({\bf q}_N/q)^{\frac
12})\rangle\,=\,\sum \limits_{\bla
\subseteq \{\CK^N\}}S_\bla ({\bf q}_N)\,
S_\bla ({\bf q}_N/q)\,= \\[0.5cm]
=\,\displaystyle{{\CV}^{-1}({\bf q}_N)
{\CV}^{-1}({\bf q}_N/q)
\det\begin{pmatrix} \displaystyle{ \frac{1-q^{(M+1)
(j+k-1)}}{1-q^{j+k-1}}}\end{pmatrix}_{1\le j,k \le N}}\,.
\end{array}
\label{gfs1}
\end{equation}
Due to Proposition~3, right-hand side of (\ref{gfs1}) is given by generating function of column strict plane
partitions (\ref{rep31}) with $L$ and $\CP$ replaced by $N$ and $\CK$, respectively:
\begin{equation}
\langle \Psi ({\bf q}_N^{-\frac
12})\mid\!\Psi (({\bf q}_N/q)^{\frac
12})\rangle\,=\,q^{-\frac{N^2}2(N-1)}
Z_{q}^{{\rm cspp}}(N, N, M)\,,
\label{spxxgf}
\end{equation}
and it coincides at $q=1$ with  the number of column strict partitions in ${{\cal B}} (N, N, M)$.

Let us proceed with the form-factor of
ferromagnetic string (\ref{spdfpxx0}).
Under the $q$-pa\-ram\-etri\-zation, the entries  (\ref{spdfpxx}) are (\ref{rep2}) with $L=N$ and $P=M-n$.
Therefore, due to Proposition~3,  we may express the form-factor as the generating function of  column strict
plane partitions though in a smaller box
${\cal B} (N, N, M-n)$:
\begin{equation}
\begin{array}{r}
\langle \Psi ({\bf q}_N^{-\frac 12})
\mid\bar\varPi_{n}\mid\!\Psi (({\bf
q}_N/q)^{\frac 12}) \rangle \,=\,
\CP_{\CK/n}({\textbf q}_N, {\textbf q}_N/q)
\\[0.4cm]
=\displaystyle{q^{n N^2} \prod_{k=1}^N
\prod_{j=1}^N
\frac{1-q^{M-n+1+j-k}}{1-q^{j+k-1}}}\,
=\,\displaystyle{q^{\frac{N^2}2(2n+1-N)}
Z^{\rm cspp}_{q}(N,N,M-n)}\,.
\end{array}
\label{spdfpxx1}
\end{equation}
The corresponding number of column strict plane partitions ({\rm AI}.4)
arises in the limit $q\to 1$:
\begin{equation}
\lim_{q\to 1}\,\langle \Psi ({\bf
q}_N^{-\frac 12})
\mid\bar\varPi_{n}\mid\!\Psi (({\bf
q}_N/q)^{\frac 12}) \rangle\,=\,{\CP}_{\CK/n} ({\textbf 1}, {\textbf 1})\,=\, A^{\rm
cspp}(N,N,M-n)\,. \label{spdfpxx3}
\end{equation}

 Due to Propositions~1 and~2, the form-factor of the domain wall creation operator, Eq.~(\ref{ratiof}), has the form in the $q$-parametrization:
\begin{equation}
\begin{array}{l}
\langle \Psi({\bf q}_{N}^{ -\frac 12})\mid \bar{\sf F}_{ n} \mid\!\Psi (({\bf q}_{N-n}/q)^{\frac 12})\rangle\,= \\[0.3cm]=\,
q^{\frac{n}2(N-n)(N-n-1)}
\sum\limits_{\bla \subseteq \{\CK^{N-n}\}} S_{\hat\bla} ({\bf q}_{N}) S_{\bla} ({\bf q}_{N-n}/q)\,
=\,\displaystyle{\frac{\det \bar{\sf T}}{{\CV}({\bf q}_{N}) {\CV}({\bf
q}_{N-n}/q)}}\,,
\end{array}
\label{rep1}
\end{equation}
where the matrix $\bar{\sf T}$ is given by (\ref{rep2}) with $L=N-n$ and $P=M$.
The partitions $\hat\bla$ and $\bla$ are defined in (\ref{field41}). Applying (\ref{rep31}), we obtain:
\begin{equation}
\langle \Psi ({\bf q}_{N}^{ -\frac 12})\mid
\bar{\sf F}_{n} \mid\!\Psi (({\bf
q}_{N-n}/q)^{ \frac 12})\rangle\,=\,q^{\frac
n2(N-n)(N-n-1)} Z_{q}^{{\rm}}(N-n, N,
{\CK})\,. \label{rep4}
\end{equation}
Therefore, the form-factor is the generating function of plane partitions confined in the box ${\cal B} (N-n, N, \CK)$. In the limit
\begin{equation}
\lim\limits_{q\to 1}\,\langle \Psi ({\bf
q}_{N}^{ -\frac 12})\mid \bar{\sf F}_{n}
\mid\!\Psi(({\bf q}_{N-n}/q)^{ \frac
12})\rangle\,=\,A(N-n, N, {\CK})\,,
\label{rep5}
\end{equation}
we obtain the correspondent MacMahon formula ({\rm AI}.2).

\section{Low temperature asymptotics}
\label{cor:sec5}

Now let us turn to the main issue of the present paper -- to the low temperature asymptotics of the correlation functions  (\ref{ratbe0}) and (\ref{field0}). We assume that our $XX0$ chain is long
enough, $M \gg 1$, while $N$ is moderate: $1\ll N \ll M$. Besides, $\be$ in (\ref{ratbe0}) and (\ref{field0}) is now inverse of the absolute temperature, $\be=\frac 1T$ (the Boltzmann constant is unity).

\subsection{Persistence
of ferromagnetic string}

For large enough $M$, the summation over the solutions to the Bethe equations in the persistence of ferromagnetic string correlation function $\CT ({\bth}^{\,\rm g}_N, n, \be)$, Eq.~(\ref{ratbe131}), is replaced by integrations over the
con\-ti\-nu\-ous variables.
Under the same assumption, the elements of $N$-tuple ${\bth}_N^{\,{\rm g}}$ of the ground state solutions (\ref{grstxx}) are such that $\cos{\theta}_{l}^{\,{\rm g}}\simeq 1$.
The approximate expression for $\CT ({\bth}^{\,\rm g}_N, n, \be)$ is of the form:
\begin{equation}
\begin{array}{r}
\CT ({\bth}^{\,\rm g}_N, n, \be)\,\simeq\,
\displaystyle{\frac{1}{{\CN}^2 ({\bth}^{\,\rm g}_N) N!}\,\int\limits_{-\pi}^{\pi}
\int\limits_{-\pi}^{\pi}\dots
\int\limits_{-\pi}^{\pi}
e^{{\be}\sum\limits_{l=1}^N
(\cos{\ta}_l- 1)} }
\\[0.3cm]
\displaystyle{\times\,\bigr|{\CP}_{\CK/n}(
e^{-i{\bth}_N}, {\bf 1})\bigr|^2 \prod_{1\leq k<l\leq N}
\bigl|e^{i{\ta}_k}-e^{i{\ta}_l}\bigr|^2\,
\frac{d\ta_1 d\ta_2\dots
d\ta_N}{(2\pi)^N}}\,.
\end{array}
\label{ratbe14}
\end{equation}
In the large $\be$ limit (low temperature limit), right-hand side of  Eq.~(\ref{ratbe14}) can be re-expressed as:
\begin{eqnarray}
&&\CT ({\bth}^{\,\rm g}_N, n, \be)\simeq
\displaystyle{\,\frac{\CA(N, n)}{\be^{\frac{N^2}{2}}}\,,\qquad {\CA}(N, n)\,\equiv\,{\CP}_{\CK/n}^2
({\textbf 1}, {\textbf 1})\,\frac{{\cal
I}_N}
{{\CN}^{2}({\bth}^{\,\rm g}_N)}}\,,
\label{ratbe141}\\[0.3cm]
&&\displaystyle{{\cal I}_N\,\equiv\,
\frac{1}{N!}\int\limits_{-\infty}^{\infty}
\!\int\limits_{-\infty}^{\infty}
\cdots\int\limits_{-\infty}^{\infty}\,e^{
-\frac{1}2\sum\limits_{l=1}^N
x^2_l}\!\!\prod_{1\leq k<l\leq N} \bigl|x_k-
x_l\bigr|^2\,\frac{d x_1 d x_2 \dots d
x_N}{(2\pi)^N} }\,. \label{ratbe15}
\end{eqnarray}
The power law decay in $\beta$ of $\CT
({\bth}^{\,\rm g}_N, n, \be)$ (\ref{ratbe141}) is
governed by the critical exponent $N^2/2$. The combinatorial factor ${\CP}^2_{\CK/n}({\textbf 1}, {\textbf 1})$ in ${\CA}(N, n)$ (\ref{ratbe141}) is, according to (\ref{spdfpxx3}), the square of the number of column strict plane partitions $A^{\rm cspp}(N,N,M-n)$.

The integral ${\cal I}_N$ (\ref{ratbe15}) is a special form of the Mehta integral related to the partition function of so-called \textit{Gaussian Unitary Ensemble} \cite{meh, selb}. Its value in terms of the gamma function \cite{magn} is known, and
it is convenient for us to put it in the exponential form:
\begin{equation}
{\cal I}_N\,=\,e^{{\varphi}_N}\,,\qquad
{{\varphi}}_N\,\equiv\,\sum\limits^N_{k=1}
\log\frac{\Gamma(k)}{(2\pi)^{1/2}}\,.
\label{ratbe161}
\end{equation}
In the considered limit the inverse of the square of the norm is equal to
\begin{equation}
\frac{1}{{\CN}^2 ({\bth}^{\,\rm g}_N)}\,
\simeq\,\frac{(2\pi)^{N(N-1)}
}{(M+1)^{N^2}}\,\prod_{1\leq r<s\leq N}
|r-s|^2\,=\,
\begin{pmatrix}\displaystyle{
\frac{2\pi}{M+1}}
\end{pmatrix}^{N^2} e^{2 {\varphi}_N}\,,
\label{ratbe142}
\end{equation}
where ${\varphi}_N$ is defined in (\ref{ratbe161}).
Taking into account (\ref{ratbe161}) and (\ref{ratbe142}), we find out that the estimate (\ref{ratbe141}) may be expressed as
\begin{equation}
\CT ({\bth}^{\,\rm g}_N, n, \be)\,\simeq\, \bigl(A^{\rm
cspp}(N, N, M-n) \bigr)^2\,
\displaystyle{e^{\Phi (N, M, \beta)}}
\label{ratbe1431}
\end{equation}
with
\begin{equation}
\Phi (N, M, \beta)\,\equiv\, N^2
\log\frac{2\pi}{M+1}-\frac{N^2}2 \log\be+ 3 {\varphi}_N \,.
\label{ratbe1432}
\end{equation}

To study the asymptotical properties of the correlation functions in question, it is appropriate to represent ${\varphi}_N$  (\ref{ratbe161}) through the Barnes $G$-function \cite{barn}:
\begin{equation}
G(z+1)\,=\,(2\pi)^{z/2}
e^{\frac{-z}{2} (z+1)-\frac{\gamma}{2} z^2}\,\prod\limits^{\infty}_{n=1}
\Bigl(1+\frac z n\Bigr)^n
e^{-z+\frac{z^2}{2n}}\,,
\label{barn1}
\end{equation}
where $\gamma$ is the Euler constant \cite{magn}.
For every non-negative integer $n$
\begin{equation}
G(n+1)\,=\,\frac{(n !)^n}{1^1\,2^2\,\ldots n^n }\,=\,
\prod^n_{k=1} \Gamma(k)\,.
\label{barn11}
\end{equation}
The theoretical aspects of the Barnes function are discussed in \cite{adam}.
The equation (\ref{barn11}) allows to
re-express $\varphi_N$ (\ref{ratbe161}):
\begin{equation}
\varphi_N\,=\,\log G(N+1) - \frac N2 \log 2\pi\,,
\label{barn13}
\end{equation}
and thus the value of the integral (\ref{ratbe15}) is given in terms of $G$-function:
\begin{equation}
{\cal I}_N\,=\, \frac{G(N+1)}{(2\pi)^{N/2}}\,.
\label{barn12}
\end{equation}

Asymptotically as $z\to\infty$:
\begin{equation}\label{barnes}
\log G(z+1)\,=\,-
\log {\cal A} + \frac z2\log 2\pi +
\begin{pmatrix}\displaystyle{\frac
{z^2}{2}-\frac {1}{12}} \end{pmatrix} \log z - \frac{3 z^2}{4} +
{\cal O} \Bigl(\frac 1z\Bigr)\,,
\end{equation}
where the constant ${\cal A}$ can be found in \cite{barn, adam}.
The behavior of
${{\varphi}}_N$ at $N\gg 1$ results from (\ref{barn13}) and (\ref{barnes}):
\begin{equation}
{{\varphi}}_N\,=\,\frac{N^2}{2} \log N\,-\,
\frac{3 N^2}{4}\,+\,{\cal O}(\log
N)\,,\qquad N\gg 1\,. \label{spdfpxx6}
\end{equation}
Hence, for the exponent (\ref{ratbe1432}) we obtain:
\begin{equation}
\Phi (N, M, \beta)\simeq N^2 \log\Bigl({\sf A} \frac{N^{3/2}}{M \be^{1/2}}\Bigr)\,,
\label{ratbe1433}
\end{equation}
where ${\sf A}$ is a constant.
In order to estimate $A^{\rm cspp} (N, N, M-n)$, we put $P=M-n$ in ({\rm AI}.4) and use (\ref{barn11}):
\begin{equation}
A^{\rm
cspp}(N,N,M-n)\,=\,
\frac{G^2(N+1)\,G(M+2-n+N)\,G(M+2-n-N)}
{G(2N+1)\,G^2(M+2-n)}\,. \label{spdfpxx5}
\end{equation}
Taking into account (\ref{barnes}), we find in the leading order:
\begin{equation}
\log A^{\rm cspp}(N, N, M-n)\,\simeq\, N^2
\log\begin{pmatrix} \displaystyle{{\sf
B}\,\frac{M-n}{N}}\end{pmatrix}\,,\quad  M-n\gg N\gg
1\,, \label{spdfpxx7}
\end{equation}
where ${\sf B}$ is a constant.
Equation (\ref{spdfpxx7}) gives us the asymptotical behavior of the number of
column strict plane partitions in a high box with a square bottom ${\cal B}(N, N, M-n)$.

Finally, taking into account (\ref{ratbe1433}) and (\ref{spdfpxx7}), we put the asymptotic estimate
(\ref{ratbe1431}) of the persistence of ferromagnetic string into the form:
\begin{equation}
\log \CT ({\bth}^{\,\rm g}_N, n, \be)\,\simeq\, N^2 \log \begin{pmatrix} \displaystyle{{\sf
C}\,\frac{(M-n)^2}{\, M (N \be)^{1/2}}}\end{pmatrix} \,,
\label{spdfpxx8}
\end{equation}
where ${\sf C}={\sf A} {\sf B}^2$.
If we assume  that $M$ and $N$ are increasing, while the temperature $T$ is decreasing, then from (\ref{spdfpxx8}) it follows that $\CT ({\bth}^{\,\rm g}_N, n, \be)$ is decreasing provided that the restriction $T <  \frac{1} {{\sf C}^{2}} \frac{N M^2}{(M-n)^4}$ holds.

\subsection{Persistence of
domain wall}

Starting with the representation (\ref{field181}) for the persistence of domain wall correlation function, we can repeat the arguments of the previous  subsection and deduce the following asymptotical expression:
\begin{equation}
{\cal F} (\widetilde{\bth}^{\,\rm g}_N, n, \be)\,\simeq \, \displaystyle{A^{2}(N-n,
N , M-N+1) }\, \displaystyle{e^{\Phi (N, M, \beta)}}\,, \label{field19}
\end{equation}
where $\Phi (N, M, \beta)$ is (\ref{ratbe1432}), and  $A(N-n, N , M-N+1)$
is the number of the plane partitions ({\rm AI}.2) in a box with
rectangular bottom ${\cal B}(N-n, N,
M-N+1)$  that may be expressed as the form-factor of the creation of the domain wall operator (\ref{rep5}).

Using ({\rm AI}.2) and (\ref{barn11}), we obtain:
\begin{eqnarray}
A(N-n,N,M-N+1)\,=\, \frac{G(N+1)\,G(N-n+1)}{G(2N-n+1)} \nonumber \\
\times\,\frac{G(M+2-n+N)\,G(M+2-N)}
{G(M+2-n)\,G(M+2)}\,.
\nonumber
\end{eqnarray}
Furthermore, we find using (\ref{barnes}):
\begin{eqnarray}
\log A(N-n, N, M-N+1)\,\simeq\, N(N-n)
\log \begin{pmatrix} \displaystyle{{\sf D}\,\frac{M-n}{2N-n}}\end{pmatrix} \,,
&&
\label{field192}\\
 M-n\gg N-n, N\gg 1\,, &&
\nonumber
\end{eqnarray}
where ${\sf D}$ is some constant. Equation (\ref{field192}) defines the asymptotical behavior of the number of plane partitions in a high box with
rectangular bottom ${\cal B}(N-n, N,
M-N+1)$. Taking into account
(\ref{ratbe1433}) and (\ref{field192}), we obtain for (\ref{field19}):
\begin{equation}
\log {\cal F} (\widetilde{\bth}^{\,\rm g}_N, n, \be)\,\simeq\, N^2 \log \begin{pmatrix} \displaystyle{{\sf
A}\,\frac{N^{3/2}}{M \be^{1/2}}}
\end{pmatrix} + 2 N(N-n) \log \begin{pmatrix} \displaystyle{ {\sf
D}\,\frac{M-n}{2N-n}}\end{pmatrix}\,.
\label{field193}
\end{equation}
Equation (\ref{field193}) enables us to state that ${\cal F} (\widetilde{\bth}^{\,\rm g}_N, n, \be)$ is decreasing with increasing $M$ and $N$ provided that the temperature $T$ is estimated analogously to the previous subsection.

\section{Discussion}
\label{cor:sec6}

In our paper we have discussed the $N$-particle thermal correlation functions of the $XX0$ Heisenberg model on a cyclic chain of a finite length. We have considered the ferromagnetic string operator $\bar\varPi_{n}$ (\ref{ratbe0}) and the domain wall creation operator $\bar{\sf F}_{n}$ (\ref{field0}). The calculations were based on the theory of the symmetric functions that allows us to express the answers in the determinantal form. The equations (\ref{spdfpxx0}) and (\ref{field4}) for the form-factors which are the basic quantities in the above correlation functions are shown to be related to the generating functions of self-avoiding random walks and boxed plane partitions. The introduced $q$-binomial determinant plays an important role in the establishing of this relation.

The estimate of the asymptotical behavior of the persistence correlation functions of the operators $\bar\varPi_{n}$ and $\bar{\sf F}_{n}$ is done for low temperatures. The problem of calculating the asymptotical
expressions leads to the calculation
of the matrix integrals of the type of (\ref{ratbe14}). These integrals  were intensively studied in various fields of theoretical physics and mathematics, \cite{loggas, wit, joh, baik, schehr}.
The low temperature approximation allows both to extract the combinatorial pre-factor and to reduce the matrix-type integrals to the partition function of the Gaussian Unitary Ensemble.
Both correlations functions have a power law decay and have the same critical exponents, but their amplitudes are different: they depend on the squared number of plane partitions contained in a box of different size. These amplitudes are observable quantities.
Expression for the corresponding exponent looks like the free energy appearing at small coupling for the large-$N$ lattice gauge theory considered in \cite{wit}. This answer should be related to the third order phase transition \cite{wit}, a possibility of which for the $XX0$ spin chain is discussed in \cite{iz5}. The results obtained in \cite{b6} for the thermal correlation functions of the $XXZ$ Heisenberg chain with the infinite coupling constant allows to argue that the third order phase transition is possible in this model as well.

\section* {Acknowledgement} This work was partially supported by RFBR (No.~13-01-00336) and by RAS program `Mathematical methods in non-linear dynamics'. We are grateful to A.~M.~Vershik for useful discussions.

\section*{Appendix~I}

Here we provide some notions concerning boxed plane partitions and their generating functions while more details can be found in \cite{macd}.

An array $(\pi _{i j})_{1\le i, j}$ of
non-negative integers that are
non-increasing as functions both of $i$ and $j$ is called {\it plane partition} $\bpi$. The integers $\pi_{i j}$ are called the {\it parts} of the plane partition, and $|\bpi| =\sum_{i, j} \pi_{i j}$ is its {\it volume}. Each plane partition has a three-dimensional diagram which can be
interpreted as a stack of unit cubes
(three-dimensional Young diagram). The
height of stack with coordinates $(i,j)$
is equal to $\pi_{i j}$. It is said that the plane partition corresponds to a box of the size $L\times N\times P$ provided that $i\leq L$, $j\leq N$ and $\pi_{i j} \leq P$ for all cubes of the Young diagram. If $\pi_{i j}>\pi_{i+1,j}$, i.e., if the parts of plane partition $\bpi$ are decaying along each column, then $\bpi $ is called the {\it column strict plane partition}.

We shall denote the box of the size $L\times N\times P$ as the set of integer lattice points:
\[{\cal B}(L, N, P)=\bigl\{(i, j, k)\in\BN^3
\bigl|\,0\leq i\leq L,\,\, 0\leq j\leq N,\,\,
0\leq k \leq P\bigr\}\,.\] An arbitrary
plane partition $\bpi$ contained in ${\cal B}(N, N, P)$ may be transferred into a column strict
plane partition $\bpi_{{\rm cspp}}$ corresponding to ${\cal
B}(N, N, P+N-1)$ by adding to $\bpi$ the $N\times N$ matrix
$$
\begin{pmatrix}
N-1 & N-1 & \cdots & N-1 \\
N-2 & N-2 & \cdots & N-2 \\
\vdots & \vdots & \ddots & \vdots \\
0 & 0 & \cdots & 0
\end{pmatrix}\,,
\nonumber
$$
which corresponds to a minimal column
strict plane partition. The volumes of the column strict plane partition and
correspondent plane partition are related:
\begin{equation}
|\bpi_{{\rm cspp}}|=|\bpi
|+\frac{N^2(N-1)}2. \nonumber
\end{equation}

The generating function of plane
partitions contained in ${\cal B}(L, N, P)$ is equal to
$$
Z_q(L, N, P)\,=\,\prod\limits_{j=1}^{L}
\prod\limits_{k=1}^{N}
\prod\limits_{i=1}^{P}
\frac{1-q^{i+j+k-1}}{1-q^{i+j+k-2}}
\,=\,\prod\limits_{j=1}^{L}
\prod\limits_{k=1}^{N}
\frac{1-q^{P+j+k-1}}{1-q^{j+k-1}}\,.
\eqno({\rm AI}.1)
$$
According to the classical MacMahon's
formula, there are
$$
A (L, N, P)\,=\,\prod\limits_{j=1}^{L}
\prod\limits_{k=1}^{N}
\prod\limits_{i=1}^{P}
\frac{i+j+k-1}{i+j+k-2}
\,=\,\prod\limits_{j=1}^{L}
\prod\limits_{k=1}^{N}
\frac{P+j+k-1}{j+k-1} \eqno({\rm AI}.2)
$$
plane partitions contained in ${\cal B}(L, N, P)$. It is clear that right-hand side of ({\rm AI}.1) tends to $A (L, N, P)$ in the limit $q\to 1$.

The generating function of the column strict plane partitions placed in ${\cal B}(N, N, P)$ is equal to
$$
Z_{q}^{{\rm cspp}}(N, N, P)\,=\,q^{\frac
{N^2}{2}(N-1)} \prod\limits_{k=1}^{N}
\prod\limits_{j=1}^{N}
\frac{1-q^{P+1+j-k}}{1-q^{j+k-1}}\,.
\eqno({\rm AI}.3)
$$
The limit $q\to 1$ gives the number $A^{{\rm cspp}}(N, N, P)$ of the
column strict partitions placed in ${\cal B}(N, N, P)$:
$$
\begin{array}{r}
\displaystyle{
A^{\rm cspp}(N, N, P)\,=\,
\prod\limits_{k=1}^{N}
\prod\limits_{j=1}^{N}
\frac{P+1+j-k}{j+k-1}} \nonumber\\
\displaystyle{
=\, \prod\limits^N_{j=1}
\frac{\Ga(j)\,\Ga(j+P+1)}{\Ga(j+N)\,
\Ga(j+P+1-N)}}\,, \nonumber
\end{array}
\eqno({\rm AI}.4)
$$
where the expression in terms of the gamma-functions \cite{magn} is
appropriate to study the asymptotics. Notice that
$$
Z_{q}^{{\rm cspp}}(N, N, P)= q^{\frac
{N^2}{2}(N-1)} Z_{q}(N, N, P-N+1)\,,
$$
and thus $A^{{\rm cspp}}(N, N,
P)=A(N, N, P-N+1)$.

\section*{Appendix~II}

The proof of Proposition~3 based on the Binet-Cauchy formula (\ref{schr}) is presented here.
First we remind some basic notions of the \textit{$q$-calculus} \cite{kac} and the \textit{symmetric functions} \cite{macd}.

The $q$-{\it
number} $[n]$ is a $q$-analogue of the
positive integer $n\in\BZ^+$,
$$
[n]\,\equiv\,\frac{1-q^n}{1-q}\,,
\eqno({\rm AII}.1)
$$
and the $q$-{\it factorial} is equal to
$$
[n]!\,\equiv\,[1]\,[2]\,\dots\,[n]\,,
\qquad [0]!\,\equiv\,1 \,. \eqno({\rm
AII}.2)
$$
The definitions ({\rm AII}.1) and ({\rm
AII}.2) allow to define the {\it q-binomial coefficient}
$\begin{bmatrix}N\\r\end{bmatrix}$:
$$
\begin{bmatrix}N\\r\end{bmatrix}\, \equiv\,
\frac{[N]\,[N-1]\,\dots\,[N-r+1]}{[r]!}\,=
\,\frac{[N]!}{[r]!\,[N-r]!}\,. \eqno({\rm
AII}.3)
$$
In the limit $q\to 1$, the $q$-binomial
coefficient $\displaystyle{
\begin{bmatrix}N\\r\end{bmatrix}}$ becomes the binomial coefficient
$\displaystyle{
\begin{pmatrix}N\\r\end{pmatrix}}$. Two
analogues of the Pascal formula exist for the $q$-binomial coefficients
$$
\begin{array}{l}
\displaystyle{
\begin{bmatrix}N\\r\end{bmatrix}\,=\,
\begin{bmatrix}N-1\\r-1\end{bmatrix}\,+\,
q^r\,
\begin{bmatrix}N-1\\r\end{bmatrix}}\,,
\\ [0.7cm] \displaystyle{
\begin{bmatrix}N\\r\end{bmatrix}\,=\,
q^{N-r}\,
\begin{bmatrix}N-1\\r-1\end{bmatrix}\,+\,
\begin{bmatrix}N-1\\r\end{bmatrix}}
\,,
\end{array}
\eqno({\rm AII}.4)
$$
where $1 \le r\le N-1$. The q-Vandermonde convolution for the $q$-binomial
coefficients has the form
$$
\displaystyle{
\begin{bmatrix}N+N^\prime\\r\end{bmatrix}\,=\,
\sum\limits_{j=0}^{\rm{min}\,(r, N)}
q^{(N-j)(r-j)}\,
\begin{bmatrix}N\\j\end{bmatrix}
\begin{bmatrix} N^\prime\\r-j\end{bmatrix}
}\,. \eqno({\rm AII}.5)
$$

The $r^{\rm th}$~order {\it elementary
symmetric function} $e_r=e_r({\bf x}_N)$
of $N$ variables, ${\bf x}_N=(x_1,
x_2, \dots, x_N)$, is defined by
$$
e_r\,\equiv\,\sum\limits_{1\leq i_1<i_2<\dots<i_r \leq N}
x_{i_1} x_{i_2} \dots x_{i_r}\,, \eqno({\rm
AII}.6)
$$
where $1\leq r \leq N$. The generating function identity can be used to define $e_r$ as the coefficients in the product:
$$
(1+t x_1) (1+t x_2) \dots (1+ t x_N) =
1+e_1 t + e_2 t^2 + \dots + e_N t^N\,.
$$
It is convenient to introduce special notations
for
$e_r$ at ${\bf x}_N={\bf q}_N$ and ${\bf
x}_N={\bf q}_N/q$ \cite{macd}:
$$
R_r(N)\,\equiv \,e_r ({\bf q}_N/q)\,=\, q^{r(r-1)/2}
\begin{bmatrix}N\\r\end{bmatrix}\,,
\qquad L_r(N)\,\equiv \, e_r ({\bf q}_N)\,=\,q^{r(r+1)/2}
\begin{bmatrix}N\\r\end{bmatrix}\,.
\eqno({\rm AII}.7)
$$

Consider the weakly decreasing partitions $\bla$, $\CP \geq \la_1\geq \la_2\geq \dots\geq \la_N\geq 0$, where $\CP\equiv P-N+1$. The {\it conjugate partition}
$\bar\bla$ is the partition with the Young diagram consisting of columns of height $\lambda_i$: $N \geq \bar\la_1\geq \bar\la_2\geq \dots\geq \bar\la_\CP\geq 0$. The corresponding strict partition ${\bar \bmu}$ is given by
${\bar \bmu} = {\bar\bla}+ {\bdl}_{\CP}$. The parts of ${\bar \bmu}$ respect:
$P \ge {\bar\mu}_{1}
> {\bar\mu}_{2} > \dots >
{\bar\mu}_{\CP} \ge 0$.
The Schur functions (\ref{sch}) related to the partition $\bla$ are expressed through the elementary symmetric functions ({\rm AII}.6) related to the conjugate partition ${\bar\bla}$  \cite{macd}:
$$
S_{\bla}({\bf x}_N)\,=\,
\det\bigl(e_{\bar\la_i -i+j}({\bf
x}_N)\bigr)_{1\le i, j\le \CP}\,. \eqno({\rm
AII}.8)
$$

In order to express the determinant of the matrix (\ref{rep2}) as the $q$-binomial determinant, Eq.~(\ref{rep30}), we will use the statement of Proposition~2 under the $q$-parametrization (\ref{rep21}):
$$
\det\bar{\sf T}\,=\, q^{\frac L2(L-1)(N-L)}
{\CV}({\bf q}_N)
{\CV}({\bf q}_L/q) \displaystyle{\sum\limits_{\bla
\subseteq \{\CP^{L}\}} S_{\hat\bla} ({\bf
q}_N) S_{\bla} ({\bf q}_L/q)}\,,
\eqno({\rm AII}.9)
$$
where the entries are:
$$
\displaystyle{ \begin{array}{ll} \bar {\sf
T}_{kj} = \displaystyle{
\frac{1-q^{(P+1)(j+k-1)}}{1-q^{j+k-1}}}\,,
& 1\le k \le
L,\,\qquad\quad 1 \le j \le N\,,\\[0.4cm]
\bar {\sf T}_{kj} = q^{j(N-k)}\,, & L+1 \le
k \le N,\,\quad 1\le j \le N\,.
\end{array}}
\eqno({\rm AII}.10)
$$
Let us denote the sum over partitions  in ({\rm AII}.9) by ${\sf\Sigma}_{\sf S}$. Applying  ({\rm
AII}.8) we bring it into the form:
$$
{\sf\Sigma}_{\sf S} \,=\,\displaystyle{\sum
\limits_{\bar\bla \subseteq \{L^{\CP}\}}
\det \bigl(e_{\bar\la_j -j+k}({\bf
q}_N)\bigr)_{1\le j, k\le \CP}\,
\det\bigl(e_{\bar\la_p -p+l}({\bf
q}_L/q)\bigr)_{1\le l, p\le \CP}} \,,
\eqno({\rm AII}.11)
$$
where summation runs over the conjugate partitions ${\bar\bla}$, $L \geq
\bar\la_1\geq \bar\la_2\geq \dots\geq
\bar\la_\CP\geq 0$. In the explicit form,
$$
\begin{array}{r}
{\sf\Sigma}_{\sf S}\, =\,\displaystyle{
\sum\limits_{\bar\bla \subseteq \{L^{\CP}\}}
\det \begin{pmatrix} L_{\bar\la_1}(N) &
L_{\bar\la_2-1}(N) & \dots &
L_{{\bar\la}_{\CP}-\CP+1}(N) \\ L_{\bar\la_1 +1}(N) &
L_{\bar\la_2}(N) & \dots &
L_{\bar\la_\CP-\CP+2}(N) \\
\vdots & \vdots & \ddots & \vdots\\
L_{\bar\la_1 +\CP-1}(N) & L_{\bar\la_2+\CP-2}(N)
& \dots & L_{\bar\la_\CP}(N)
\end{pmatrix}} \\ [1.3cm]
\times\,\displaystyle{ \det\begin{pmatrix}
R_{\bar\la_1}(L) & R_{\bar\la_1+1}(L) & \dots &
R_{\bar\la_1+\CP-1}(L) \\ R_{\bar\la_2 -1}(L) &
R_{\bar\la_2}(L) & \dots &
R_{\bar\la_2+\CP-2}(L)\\
\vdots & \vdots & \ddots & \vdots\\
R_{\bar\la_\CP -\CP+1}(L) &
R_{\bar\la_\CP-\CP+2}(L) & \dots &
R_{\bar\la_\CP}(L)
\end{pmatrix}}
\,,
\end{array}
\eqno({\rm AII}.12)
$$
where the notations ({\rm AII}.7) are
taken into account. By definition, $R_0=L_0=1$,
while both $R_r$ and $L_r$
are equal zero for $r>N$. The  Binet-Cauchy formula allows to express
({\rm AII}.12) as the determinant of the $\CP\times\CP$
matrix:
$$
{\sf\Sigma}_{\sf S} = \displaystyle{
\det\begin{pmatrix}
\sum\limits_{i=0}^{L} L_{i}(N) R_{i}(L) &
\sum\limits_{i=0}^{L} L_{i}(N) R_{i+1}(L)
  & \dots & \sum\limits_{i=0}^{L}
  L_{i}(N) R_{i+\CP-1}(L) \\
\sum\limits_{i=0}^{L} L_{i+1}(N) R_{i}(L) &
\sum\limits_{i=0}^{L} L_{i}(N) R_{i}(L) & \dots
&\sum\limits_{i=0}^{L}  L_{i}(N) R_{i+\CP-2}(L)
\\ \cdots & \cdots & \ddots & \cdots\\
 \sum\limits_{i=0}^{L} L_{i+\CP-1}(N) R_{i}(L)
 & \sum\limits_{i=0}^{L}
 L_{i+\CP-2}(N) R_{i}(L) & \dots &
\sum\limits_{i=0}^{L} L_{i}(N) R_{i}(L)
\end{pmatrix}
} . \eqno({\rm AII}.13)
$$
Calculating the entries in ({\rm AII}.13)  by the $q$-Vandermonde convolution ({\rm AII}.5),
$$
\begin{array}{l}
\displaystyle{ \sum \limits_{i=0}^{L}
L_{i+s}(N) R_{i}(L)\,=\,q^{\frac{s}2(s+1)}\,
\begin{bmatrix} N+L
\\N-s\end{bmatrix}}\,,
 \\ [0.5cm]
\displaystyle{\sum \limits_{i=0}^{L} L_{i}(N)
R_{i+s}(L)\,=\,q^{\frac{s}2(s-1)}\,
\begin{bmatrix} N+L
\\N+s\end{bmatrix}} \,,
\end{array}
\eqno({\rm AII}.14)
$$
where $0\le s \le \CP-1$,
we express ${\sf\Sigma}_{\sf S}$ as the determinant of the matrix
with the entries given by the $q$-binomials
$$
{\sf\Sigma}_{\sf S} = \displaystyle{
 \det\left(q^{\frac{i-j}{2}(i-j+1)}
 \begin{bmatrix}
 L+N \\ N-i+j \end{bmatrix} \right)_{1\le i, j\le
 \CP}}\,.
\eqno({\rm AII}.15)
$$

To bring this determinant to the $q$-binomial form, Eq.~(\ref{bd0}), we will use the Pascal formulas ({\rm
AII}.4). As a first step, we combine the
rows in ({\rm AII}.15) in the following way: the $\CP^{\rm th}$ is multiplied by $q^{N+1}$
and the $(\CP-1)^{\rm th}$ is added to it,
the $(\CP-1)^{\rm th}$ is multiplied by
$q^{N+1}$ and the $(\CP-2)^{\rm th}$ is
added to it; $\dots$; the $2^{\rm nd}$ is
multiplied by $q^{N+1}$ and the $1^{\rm
st}$ is added to it. The second step is
concerned with the matrix obtained: the
$\CP^{\rm th}$ row is multiplied by $q^{N+2}$ and the $(\CP-1)^{\rm th}$ one is added to it, the $(\CP-1)^{\rm th}$ row is multiplied by $q^{N+2}$ and the $(\CP-2)^{\rm th}$ one is added to it; $\dots$; the $3^{\rm rd}$ is
multiplied by $q^{N+2}$ and the $2^{\rm
nd}$ is added to it. After $\CP-1$ steps, compensating the obtained factor by
$$
q^{\sum\limits_{k=1}^{\CP-1}
k(k-N-\CP)}\,=\,q^{\frac{N}2(1-\CP)\CP +
\frac{\CP}6(1-{\CP}^2)}\,,
$$
we obtain the desired  answer:
$$
{\sf\Sigma}_{\sf S}\,=\,
q^{-\frac
{N}{2}({\CP}-1) {\CP}}
\begin{pmatrix}
L+N, & L+N+1, & \dots & L+N+{\CP}-1\\
L, & L+1, & \dots & L+{\CP}-1
\end{pmatrix}_q\,,
\eqno({\rm AII}.16)
$$
what justifies (\ref{rep30}).

To calculate the $q$-binomial determinant in ({\rm AII}.16), we drop
$[N+L]$ out of the $1^{\rm st}$ row,
$[N+L+1]$ out of the $2^{\rm nd}$ row; $\dots$; $[N+L +\CP-1]$ out of the last one. We drop $\frac1{[L]}$ out of the $1^{\rm st}$ column, $\frac1{[L+1]}$ out of the $2^{\rm nd}$ column; $\dots$; $\frac1{[L
+\CP-1]}$ out of the last one. This
operation is repeated until all the entries of the first column become unities, and we obtain
the answer in the form:
$$
{\sf\Sigma}_{\sf S}\,=\,
q^{-\frac
{N}{2}({\CP}-1) {\CP}}\,
\displaystyle{\prod \limits_{j=1}^{L} \prod
\limits_{k=1}^{\CP}
\frac{[N+j+k-1]}{[j+k-1]}}\,\times\,
\begin{pmatrix}
N, & N+1, & \dots & N+{\CP}-1\\
0, & 1, & \dots & {\CP}-1
\end{pmatrix}_q\,.
\eqno({\rm AII}.17)
$$
After the standard transformations, the
determinant in ({\rm AII}.17) acquires the form:
$$
q^{\frac
{N}{2}({\CP}-1) {\CP}} \det(Q_{a b})_{1\le
a, b\le {\CP}}\,,\qquad Q_{a b}\,=\,\Biggl\{
 \begin{array}{ll}
 q^{(a-1)(a-b)} \begin{bmatrix} N\\ b-a
 \end{bmatrix}\,,& a\le b  \\
  0\,, & a > b \end{array} \Biggr.\,.
$$
Since $\det Q=1$, the $q$-binomial determinant in ({\rm AII}.17) is equal to $q^{\frac{N}2(\CP-1)\CP}$, and ${\sf\Sigma}_{\sf S}$ becomes the double product thus justifying the double product in (\ref{rep31}).

The double product in ({\rm AII}.17)
coincides with the generating function of boxed plane partitions $Z_q(L, {\CP}, N)$
(see ({\rm AI}.1)):
$$
\displaystyle{\prod \limits_{j=1}^{L} \prod
\limits_{k=1}^{\CP}
\frac{[N+j+k-1]}{[j+k-1]}\,\equiv\,\prod
\limits_{j=1}^{L} \prod \limits_{k=1}^{\CP}
\frac{1-q^{N+j+k-1}}{1-q^{j+k-1}}\,=\,
Z_q(L, {\CP}, N)}\,, \eqno({\rm AII}.18)
$$
Since the generating function of the boxed plane partitions is invariant under permutations of the box sides $L, {\CP}, N $, then
$$
\displaystyle{ {\sf\Sigma}_{\sf S}\,=\,Z_q
(L, N, {\CP}) }\,. \eqno({\rm AII}.19)
$$
Substituting ${\sf\Sigma}_{\sf S}$ ({\rm
AII}.19) into ({\rm AII}.9), we obtain the second statement expressed by
(\ref{rep31}).~$\Box$

\end{document}